\documentclass[12pt]{article}

\usepackage{amsmath,amssymb}
\usepackage[dvips,dvipdfm]{graphicx}
\usepackage{bm}

\newlength{\extraspace}
\newlength{\extraspaces}
\catcode`\@=11
\def\numberbysection{\@addtoreset{equation}{section}
\def\theequation{\arabic{section}.\arabic{equation}}}
\newcommand{\newsection}[1]{
\vspace{7mm}
\pagebreak[3]
\addtocounter{section}{1}
\setcounter{equation}{0}
\setcounter{subsection}{0}
\setcounter{footnote}{0}
\begin{center}
{\large \textbf{\thesection. #1}}
\end{center}
\nopagebreak
\medskip
\nopagebreak
\hspace{3mm}}
\newcommand{\nonu}{\nonumber \\[.5mm]}
\newcommand{\A}{&\!\!\!}

\newcommand{\VEV}[1]{\left\langle {#1} \right\rangle}

\begin{document}
\addtolength{\baselineskip}{.7mm}
\thispagestyle{empty}
\begin{flushright}
STUPP--07--193 \\
October, 2007
\end{flushright}
\vspace{15mm}
\begin{center}
{\Large \textbf{Chiral Symmetry Breaking in Brane Models 
}} \\[15mm]
\textsc{Norio Horigome},${}^{\rm a}$\footnote{
\texttt{e-mail: horigome@krishna.th.phy.saitama-u.ac.jp}}
\hspace{1mm}
\textsc{Madoka Nishimura}${}^{\rm b}$
\hspace{0mm} and \hspace{0mm}
\textsc{Yoshiaki Tanii}${}^{\rm a}$\footnote{
\texttt{e-mail: tanii@phy.saitama-u.ac.jp}} \\[7mm]
${}^{\rm a}$\textit{Division of Material Science \\ 
Graduate School of Science and Engineering \\
Saitama University, Saitama 338-8570, Japan} \\[3mm]
${}^{\rm b}$\textit{Department of Community Service and Science \\
Tohoku University of Community Service and Science \\
Iimoriyama 3-5-1, Sakata 998-8580, Japan} \\[10mm]
\textbf{Abstract}\\[7mm]
{\parbox{13cm}{\hspace{5mm}
We discuss the chiral symmetry breaking in general intersecting 
D$q$/D$p$ brane models consisting of $N_c$ D$q$-branes and a 
single D$p$-brane with an $s$-dimensional intersection. 
There exists a QCD-like theory localized at the intersection 
and the D$q$/D$p$ model gives a holographic description of it. 
The rotational symmetry of directions transverse to both of the 
D$q$ and D$p$-branes can be identified with a chiral symmetry, 
which is non-Abelian for certain cases. 
The asymptotic distance between the D$q$-branes and 
the D$p$-brane corresponds to a quark mass. 
By studying the probe D$p$-brane dynamics in a D$q$-brane background 
in the near horizon and large $N_c$ limit we find that the chiral 
symmetry is spontaneously broken and there appear 
(pseudo-)Nambu-Goldstone bosons. 
We also discuss the models at finite temperature. 
}}
\end{center}
\vfill
\newpage
\setcounter{section}{0}
\setcounter{equation}{0}
\numberbysection
%
%
\newsection{Introduction}
%
%
The AdS/CFT correspondence 
\cite{Maldacena:1997re,Gubser:1998bc,Witten:1998qj} 
(see \cite{Aharony:1999ti} for a review) provides a new 
non-perturbative approach to strongly coupled 
gauge theories. This duality relates a string theory in 
($d+1$)-dimensional anti de Sitter spacetime (times a compact 
space) to a $d$-dimensional conformal field theory. 
The AdS/CFT correspondence can be extended to the string/gauge 
duality, which is a generalization to non-conformal and 
non-supersymmetric theories. The string/gauge duality also 
provides us with a useful tool for the analysis of low energy 
behaviors of QCD such as the confinement and the spontaneous 
chiral symmetry breaking.
This approach is often called the holographic QCD
\cite{Karch:2002sh, Sakai:2003wu, Babington:2003vm,
Kruczenski:2003uq,Evans:2004ia,Ghoroku:2004sp,Bak:2004nt,Rodriguez:2005jr,
Sakai:2004cn,Sakai:2005yt} (and references therein). 
\par
%
%
One of the most interesting phenomena of the low energy QCD
is the spontaneous breaking of the chiral symmetry.
In the holographic approach the chiral symmetry can be realized 
in two different ways. 
%
In this approach one introduces $N_{c}$ color D$q$-branes 
and $N_{f}$ flavor D$p$-branes. 
The U($N_c$) gauge field on the D$q$-branes represents a gluon 
field of a QCD-like theory. Open strings connecting the D$q$-branes 
and the D$p$-branes represent quarks in the fundamental 
representation of U($N_c$). 
When these brane configurations have directions transverse to 
both of the D$q$ and D$p$-branes, a rotational symmetry in these 
directions can be understood as a chiral symmetry of the dual 
gauge theories in certain cases \cite{Babington:2003vm,
Kruczenski:2003uq,Evans:2004ia,Ghoroku:2004sp,Bak:2004nt,Rodriguez:2005jr}. 
One can separate color branes and flavor branes in such directions. 
%
%
The asymptotic distance between these branes is identified with 
a quark mass. So one can study the spontaneous chiral symmetry 
breaking starting from a theory with a non-vanishing quark mass 
and taking the massless limit. 
So far only 
the Abelian chiral symmetry U(1)${}_V$ $\times$ U(1)${}_A$ 
is considered in this approach. 
\par
Alternatively, the chiral symmetry can be realized as a gauge 
symmetry on the flavor branes. 
When D$p$-$\overline{\text{D}p}$-brane pairs are used as flavor 
branes, one can obtain a non-Abelian $\text{U}(N_f)_L$ $\times$ 
$\text{U}(N_f)_R$ chiral symmetry 
\cite{Sakai:2004cn,Sakai:2005yt,Antonyan:2006vw,Gao:2006up,
Antonyan:2006qy,Basu:2006eb,Antonyan:2006pg,Gepner:2006qy}. 
These configurations of physical interest often do not have 
directions transverse to both of the color and flavor branes. 
Therefore, it is not obvious how to introduce a quark mass 
in these models. 
For work toward an introduction of a quark mass in this type 
of models and on related issues see refs.\ \cite{Casero:2007ae,
Hashimoto:2007fa,Evans:2007jr,Bergman:2007pm,Dhar:2007bz}. 
In both of these two approaches, the spontaneous breaking 
of the chiral symmetry is closely related to the configurations 
of the probe branes in the background geometry. 
\par
The chiral symmetry breaking was also discussed at finite temperature 
\cite{Babington:2003vm,Kruczenski:2003uq,Ghoroku:2005tf,
Aharony:2006da,Parnachev:2006dn,Mateos:2006nu,Peeters:2006iu,
Mateos:2007vn} and at finite chemical potential 
\cite{Kim:2006gp, Horigome:2006xu,
Parnachev:2006ev,Nakamura:2006xk,Kobayashi:2006sb,Mateos:2007vc}.
The temperature $T$ is related to a period $\delta t_E$ of the 
S$^{1}$ compactified Euclidean time coordinate as 
$T = 1/ \delta t_{E}$. The chemical potential $\mu$ is introduced 
as a non-vanishing asymptotic value of the time component of the 
gauge field on the probe brane $A_{0} \sim \mu$. 
One can study a chiral phase transition 
and obtain a phase diagram of the QCD-like theories.
\par
%
%
The purpose of the present paper is to study the chiral symmetry 
breaking in general intersecting D$q$/D$p$ brane systems consisting 
of $N_{c}$ color D$q$-branes and a single probe D$p$-brane with 
an $s$-dimensional intersection. 
They are holographic duals of QCD-like theories in 
$(s+1)$-dimensional spacetime QCD${}_{s+1}$ in certain cases. 
As in refs.\ \cite{Babington:2003vm,Kruczenski:2003uq} these models 
can have directions transverse to both of the D$q$ and D$p$-branes. 
A rotational symmetry of these directions can be interpreted as 
a chiral symmetry in certain cases. 
This symmetry can be non-Abelian in contrast to the models in 
refs.\ \cite{Babington:2003vm,Kruczenski:2003uq}. 
We can separate the D$q$-branes and the D$p$-brane in these 
transverse directions and break the rotational symmetry. 
In the holographic description this deformation makes quarks 
on the intersection massive and leads to an explicit chiral 
symmetry breaking. 
In the near horizon limit and the large $N_c$ limit we can treat 
the D$q$-branes as a background geometry and the D$p$-brane as 
a probe which does not affect this background.
We discuss the chiral symmetry breaking by analyzing the 
D$p$-brane dynamics in the D$q$-brane background geometry.
\par
%
%
The organization of this paper is as follows.
In sect.\ 2 we study the low energy spectrum at an $s$-dimensional 
intersection of the D$q$/D$p$ brane system. 
In general dual theories are defect field theories 
\cite{DeWolfe:2001pq,Constable:2002xt}. We are interested 
in field theories without defects. There are systems 
corresponding to QCD-like theories at the intersection. 
In particular, the D$2$/D$4$ model with $s=1$, the D$3$/D$5$ model 
with $s=2$ and the D$4$/D$6$ model with $s=3$ correspond to 
QCD${}_2$, QCD${}_3$ and QCD${}_4$, respectively. 
For certain $(q,p,s)$ the rotational symmetry of the 
transverse directions can be understood as a chiral 
symmetry in the QCD-like theories.
This chiral symmetry is non-Abelian 
SU(2)${}_L$ $\times$ SU(2)${}_R$ for QCD${}_2$. 
\par
In sect.\ 3 we study the chiral symmetry breaking in the 
QCD-like theories by using a supergravity analysis.
The near horizon limit and the large $N_c$ limit $N_{c} \gg 1$ 
allow us to study the probe D$p$-brane dynamics in the D$q$-brane 
background. We find that a D$p$-brane embedding breaks the 
rotational symmetry of the transverse directions. 
This corresponds to a chiral symmetry breaking in the QCD-like 
theories. The quark mass $m_{q}$ and the quark condensate 
$\VEV{\bar{\psi}\psi}$ can be read from the asymptotic behavior 
of the D$p$-brane embedding. There is a non-zero value of the quark 
condensate $\VEV{\bar{\psi}\psi}$ even for the massless quark limit.
This leads to a spontaneous chiral symmetry breaking
in the QCD-like theories. 
\par
In sect.\ 4 we consider fluctuations of the D$p$-brane 
around the vacuum embedding discussed in sect.\ 3. 
For $m_{q}=0$ there appear $(8-q-p+s)$ massless scalar bosons, 
which can be understood as the Nambu-Goldstone (NG) bosons 
associated with the spontaneous symmetry breaking.
For a non-zero but small quark mass there appear pseudo-NG bosons.
We show that these pseudo-NG bosons satisfy the 
Gell-Mann-Oakes-Renner (GMOR) relation \cite{Gell-Mann:1968rz}. 
The effective action of the fluctuations is obtained at quartic order.
\par
%
%
In sect.\ 5 we discuss the theories at finite temperature. 
We study the probe D$p$-brane dynamics in the Euclidean D$q$ 
background. The D$p$-brane embedding breaks the rotational symmetry 
as in the zero temperature case. Then the chiral symmetry is also 
broken at finite temperature. 
We find that the quark condensate vanishes and the chiral 
symmetry is restored only in the high temperature limit. 
We also study the models with $s=q$. 
%
We conclude in sect.\ 6. 

%
\newsection{General setup}
We consider an intersecting brane system consisting of 
$N_{c}$ color D$q$-branes and a single probe D$p$-brane 
\begin{equation}
\begin{tabular}{rcccccccccccc}
& $x^{0}$ & $\cdots$ & $x^{s}$ & $x^{s+1}$ & $\cdots$ 
& $x^{q}$ & $x^{q+1}$ & $\cdots$ & $x^{q+p-s}$ & $x^{q+p-s+1}$ 
& $\cdots$ & $x^{9}$\\ \hline
$N_c$ D$q$ & $\circ$ & $\cdots$ & $\circ$ & $\circ$ & $\cdots$ 
& $\circ$ & $-$ & $\cdots$ & $-$ & $-$ & $\cdots$ & $-$ \\
D$p$ & $\circ$ & $\cdots$ & $\circ$ & $-$ & $\cdots$ & $-$ 
& $\circ$ & $\cdots$ & $\circ$ & $-$ & $\cdots$ & $-$ \\[-3mm]
\label{config_dq/dp}
\end{tabular}
\end{equation}
with $x^q$ being a coordinate of S${}^1$. It has an $s$-dimensional 
intersection in the directions $x^1, \cdots, x^s$. 
The configuration (\ref{config_dq/dp}) is a T-dual 
of D$s'$/D9 system with $s'=9-(q+p-2s) \ge s$. 
Following ref.\ \cite{Antonyan:2006pg} we call it 
a \textit{transverse intersection} if $s'=s$ ($q+p-s=9$) and 
a \textit{non-transverse intersection} if $s' > s$ ($q+p-s<9$). 
Non-transverse intersections have directions transverse to both 
of the D$q$-branes and the D$p$-brane, while transverse 
intersections do not. 
\par
%
The configuration (\ref{config_dq/dp}) has the following symmetries. 
The gauge symmetry of this system is U($N_{c}$) $\times$ U(1). 
The U(1) gauge symmetry on the D$p$-brane is regarded as 
a global symmetry (baryon number symmetry) of 
an ($s+1$)-dimensional field theory at the intersection.
The ten-dimensional Lorentz symmetry SO($1,9$) is broken 
to its subgroup by the configuration (\ref{config_dq/dp}). 
Therefore the global symmetry preserved at the intersection contains 
\begin{equation}
\text{SO}(1,s) \times \text{SO}(9-q-p+s) \times \text{U}(1), 
\label{symmetry_dq/dp}
\end{equation}
where SO($1,s$) is the Lorentz symmetry at the intersection 
and SO($9-q-p+s$) is the rotational symmetry in the 
directions $x^{q+p-s+1}, \cdots, x^9$. 
\par
The spectrum of the theory localized at the intersection 
is as follows. Massless fields generated by $q$-$q$ strings 
(open strings having both ends on the D$q$-branes) are a gauge 
field $A_{\mu}$ ($\mu = 0,1,\cdots, s$), scalar fields 
$\Phi^{i}$ ($i = s+1, \cdots , 9$) and fermionic fields $S$. 
Imposing the periodic boundary condition for the bosonic fields 
and the anti-periodic one for the fermionic fields along the 
compact $x^{q}$ direction, the fermions become massive at zero 
mode and supersymmetry 
is explicitly broken at low energy. Then the scalars acquire mass 
at one-loop level. Thus only the gauge field $A_{\mu}$ is massless 
at low energy. This gives a pure U($N_c$) gauge theory.
\par
To study the lowest modes generated by $q$-$p$ strings 
(open strings connecting the D$q$-branes and the D$p$-brane), 
we note the zero-point energy in the R sector and 
the NS sector \cite{Polchinski:1998rr}
\begin{equation}
a^{\text{R}} = 0, \qquad 
a^{\text{NS}} = \frac{\text{\#ND}-4}{8}, 
\label{zeropoint}
\end{equation}
where $\text{\#ND} = q+p-2s = 9-s'$ is the number of spatial 
coordinates of open strings which have the Neumann boundary 
condition for one end and the Dirichlet one for the other end. 
The lowest modes generated by $q$-$p$ strings in the NS sector 
are massive for $\text{\#ND} > 4$ ($\text{\#ND}=6$, 8), 
massless for $\text{\#ND} = 4$ 
and tachyonic for $\text{\#ND} < 4$ ($\text{\#ND}=0$, 2).
We do not consider the tachyonic case $\text{\#ND} < 4$. 
When $\text{\#ND} \ge 4$, the lowest modes from the NS sector 
are massive (by loop effects for $\text{\#ND} = 4$) and are 
decoupled at low energy. There are only massless fermions 
from the R sector. 
They belong to representations of the Clifford algebra for 
the NN and DD directions. 
These fermions belong to the fundamental 
representation of U($N_c$) and are called ``quarks''. 
\par
In general the D$q$/D$p$ configuration (\ref{config_dq/dp}) is 
dual to a defect field theory \cite{DeWolfe:2001pq,Constable:2002xt}.
We only consider the case $s+1=q$, which corresponds to a theory 
without defects. 
We are interested in non-transverse intersections satisfying $s' > s$, 
which implies $s < 9 - \text{\#ND}$. 
Possible cases are $s=1$, 2, 3, 4 for $\text{\#ND}=4$ ($p=q+2$) 
and $s=1$, 2 for $\text{\#ND}=6$ ($p=q+4$). 
The configurations with $\text{\#ND}=6$ do not preserve supersymmetry 
and are most likely unstable. 
%
We further restrict ourselves to the cases $s=1$, 2, 3 since we are 
especially interested in theories in four and lower dimensions. 
%
%
%
To summarize, we consider the D$q$/D$(q+2)$ configurations 
for $q=2$, 3, 4 compactified on $x^q$ shown in Table \ref{conf_qcd_s+1}. 
The effective theory on the intersection at low energy 
is an $(s+1)$-dimensional non-supersymmetric U($N_{c}$) gauge theory 
with quarks in the fundamental representation. 
We call this theory ``QCD$_{s+1}$'' for the sake of convenience. 
\begin{table}[!t]
\begin{center}
\begin{tabular}{|c|c|cccccccccc|} \hline
& & $0$ & $1$ & $2$ & $3$ & $4$ & 
        $5$ & $6$ & $7$ & $8$ & $9$  \\ \hline
color & D$2$ & $\circ$ & $\circ$ & $\circ$ & $-$ & $-$ & 
        $-$ & $-$ & $-$ & $-$ & $-$  \\ 
probe & D$4$ & $\circ$ & $\circ$ & $-$ & $\circ$ & $\circ$ & 
        $\circ$ & $-$ & $-$ & $-$ & $-$ \\
\hline
color & D$3$ & $\circ$ & $\circ$ & $\circ$ & $\circ$ & $-$ & 
        $-$ & $-$ & $-$ & $-$ & $-$ \\ 
probe & D$5$ & $\circ$ & $\circ$ & $\circ$ & $-$ & $\circ$ & 
        $\circ$ & $\circ$ & $-$ & $-$ & $-$  \\
\hline
color & D$4$ & $\circ$ & $\circ$ & $\circ$ & $\circ$ & $\circ$ & 
        $-$ & $-$ & $-$ & $-$ & $-$ \\ 
probe & D$6$ & $\circ$ & $\circ$ & $\circ$ & $\circ$ & $-$ & 
        $\circ$ & $\circ$ & $\circ$ & $-$ & $-$  \\
\hline
\end{tabular}
\caption[]{The D$q$/D$(q+2)$ brane configurations with $\text{\#ND}=4$.
From top to bottom these are dual to $\text{QCD}_{2}$, $\text{QCD}_{3}$ 
and $\text{QCD}_{4}$, respectively.
}
\label{conf_qcd_s+1}
\end{center}
\end{table}
\par
Since these configurations are non-transverse intersections, 
there are directions transverse to both of the D$q$-branes and 
the D$p$-brane. 
In refs.\ \cite{Babington:2003vm, Kruczenski:2003uq} 
the rotational symmetry SO($9-q-p+s$) 
of such directions is interpreted as a chiral symmetry 
in the dual gauge theory for certain sets of $(q,p,s)$. 
When the D$q$-branes and the D$p$-brane are separated along these 
directions, quarks on the intersection become massive and 
the chiral symmetry is explicitly broken. 
As we will see in sect.\ 3, only when 
$a^{\text{NS}} = 0$ ($\text{\#ND}=4$) and $p-s-2 > 0$, which are 
satisfied for the configurations in Table \ref{conf_qcd_s+1}, 
an equation for probe brane embeddings derived from the DBI 
action has a solution for which the distance 
between the color branes and the probe brane asymptotically 
approaches a constant value. This distance is interpreted as 
a quark mass. 
%
%
\par
We can explicitly write down the symmetry (\ref{symmetry_dq/dp}) 
for the configurations in Table \ref{conf_qcd_s+1}. 
In the D$2$/D$4$ model, which has a one-dimensional intersection 
and is dual to QCD$_{2}$, we can identify the SO(4)$_{6789}$ 
rotational symmetry in the $x^6$, $x^7$, $x^8$, $x^9$ directions 
with an $\text{SU(2)}_{L} \times \text{SU(2)}_{R}$ chiral 
symmetry of quarks. 
Indeed, the GSO projection in the R sector of open strings 
requires that the chiralities of SO($1,1$)${}_{01}$ and 
SO(4)${}_{6789}$ are correlated. 
Left-handed (right-handed) quarks of SO($1,1$)${}_{01}$ have 
the positive (negative) chirality of SO(4)${}_{6789}$ and 
transform as $({\bf 2},{\bf 1})$ ($({\bf 1}, {\bf 2})$) under 
$\text{SU}(2)_{L} \times \text{SU}(2)_{R}$. 
The gauge symmetry U(1) on the probe brane acts on quarks 
as a baryon number symmetry U(1)${}_V$. 
Therefore the global symmetry (apart from the Lorentz symmetry) 
of QCD$_{2}$ at the intersection is 
\begin{equation}
\text{SO}(4)_{6789} \times \text{U}(1)
\sim \text{SU}(2)_{L} \times \text{SU}(2)_{R} \times \text{U}(1)_{V}. 
\end{equation}
Thus we can realize a non-Abelian chiral symmetry in a holographic 
model of this type, although spacetime is two-dimensional. 
\par
In the D$3$/D$5$ model, which has a two-dimensional intersection
and is dual to QCD$_{3}$, we can identify the SO(3)$_{789}$ 
rotational symmetry in the $x^7$, $x^8$, $x^9$ directions with 
an $\text{SU}(2)$ symmetry of QCD$_{3}$. Then the global symmetry 
of QCD$_{3}$ at the intersection is 
\begin{equation}
\text{SO}(3)_{789} \times \text{U}(1)
    \sim \text{SU}(2) \times \text{U}(1).
\label{symmetry_d3/d5}
\end{equation}
Quarks transform as {\bf 2} under SU(2). 
Note that there is no chirality in QCD${}_3$ and therefore 
the symmetry (\ref{symmetry_d3/d5}) is not a chiral symmetry. 
\par
%
Finally, in the D$4$/D$6$ model, which has a three-dimensional 
intersection and is dual to QCD$_{4}$, we can identify the 
SO(2)$_{89}$ rotational symmetry in the $x^8$, $x^9$ directions 
with an axial $\text{U}(1)_A$ symmetry of QCD$_{4}$ as discussed 
in refs.\ \cite{Babington:2003vm, Kruczenski:2003uq}. 
The global symmetry of QCD$_{4}$ at the intersection is 
\begin{equation}
\text{SO}(2)_{89} \times \text{U}(1)
    \sim \text{U}(1)_A \times \text{U}(1)_V.
\end{equation}
\par

%
\newsection{Chiral symmetry breaking from supergravity analysis}
The dynamics of a strongly coupled large $N_{c}$ gauge theory 
can be analyzed by supergravity. 
We study the chiral symmetry breaking in this section. 
The near horizon limit and the large $N_c$ limit $N_{c} \gg 1$ 
allow us to treat the D$q$-branes as a background geometry 
and the D$p$-brane as a probe which does not affect this background.
We will find that the D$p$-brane embedding breaks 
the SO($9-q-p+s$) rotational symmetry in the directions 
transverse to both of the branes. 
This can be interpreted as the chiral symmetry breaking 
in QCD$_{2}$ and QCD$_{4}$. 
We will calculate the quark condensate and find a non-zero value 
even in the massless quark limit. 
Although we are most interested in the configurations in 
Table \ref{conf_qcd_s+1}, we will give formulae for the configuration 
(\ref{config_dq/dp}) with general $q$, $p$, $s$. 

%
\subsection{D$q$-brane background}

The near horizon geometry of S$^1$ compactified 
$N_c$ D$q$-branes is \cite{Itzhaki:1998dd}
\begin{gather}
ds^2 = \left( \frac{U}{R}\right)^{\frac{7-q}{2}} \! 
    \left( - dt^2 + \sum_{i=1}^{q-1} \left( dx^{i} \right)^2 
        + f(U) (dx^{q})^2\right) 
    + \left( \frac{R}{U}\right)^{\frac{7-q}{2}} \!
        \left( \frac{dU^2}{f(U)} + U^2 d\Omega_{8-q}^{2} \right), \notag \\
f(U) = 1 - \left( \frac{U_{KK}}{U} \right)^{7-q}, \qquad
R^{7-q} = (4\pi)^{\frac{5-q}{2}} \, \Gamma \left( \tfrac{7-q}{2} \right) 
            g_s N_c \; \ell_s^{\;7-q},
\label{dq-bg_1}
\end{gather}
where $d\Omega_{8-q}^{2}$, $g_{s}$ and $\ell_{s}$ are the metric 
of a unit S$^{8-q}$, the string coupling and the string length, 
respectively.
$x^q$ is a coordinate of S$^{1}$ and its period is denoted as 
$\delta x^q = 2\pi / M_{KK}$. 
$M_{KK}$ is a compactification scale. 
To avoid a conical singularity at $U=U_{KK}$ in the $U$-$x^q$ plane 
the period must be related to a constant $U_{KK}$ as 
\begin{equation}
\delta x^q
= \frac{4 \pi \; R^{\frac{7-q}{2}}}
    { (7-q) \; U_{KK}^{\frac{5-q}{2}}}.
\label{delta_x_q}
\end{equation}
The dilaton field and the Ramond-Ramond (RR) flux are given by
\begin{equation}
e^{\phi} = g_{s} \left( \frac{R}{U} \right)^{\frac{(7-q)(3-q)}{4}}, 
\qquad 
F_{8-q} = \frac{N_{c}}{V_{8-q}} \epsilon_{8-q},
\label{dilaton_flux}
\end{equation}
where $\epsilon_{8-q}$ and $V_{8-q}$ are 
the volume form and the volume of a unit S$^{8-q}$. 
\par
The relations between the parameters in the gauge theory 
and those in the string theory are
\begin{equation}
g_{q+1}^{2} = (2\pi)^{q-2} g_s \, \ell_s^{\;q-3}, \qquad 
M_{KK} = \frac{7-q}{2 \, (4\pi)^{\frac{5-q}{4}} \, 
        \Gamma(\tfrac{7-q}{2})^{\frac{1}{2}}} 
        \frac{U_{KK}^{\frac{5-q}{2}}}
            {(g_{s} N_{c})^{\frac{1}{2}} \, \ell_s^{\;\frac{7-q}{2}}}, 
\label{g_m_kk}
\end{equation}
where $g_{q+1}$ is the $(q+1)$-dimensional gauge coupling. 
The $(q+1)$-dimensional 't~Hooft coupling is defined as 
\begin{equation}
\lambda_{q+1} = \frac{g_{q+1}^{2} N_{c}}{(2\pi)^{q-2}}. 
\end{equation}
Note that the supergravity description is 
valid for \cite{Itzhaki:1998dd, Antonyan:2006pg} 
\begin{equation}
1 \ll \lambda_{q+1} \left( \frac{U_{KK}}{\ell_s^2} \right)^{q-3} 
\ll N_{c}^{\frac{4}{7-q}}. 
\end{equation}
\par
We introduce isotropic coordinates in the directions 
$(U, \Omega_{8-q})$ to simplify the following analysis.
Introducing a new radial coordinate $\rho$ defined by
\begin{equation}
U = \left( \rho^{\frac{7-q}{2}} 
    + \frac{U_{KK}^{7-q}}{4 \rho^{\frac{7-q}{2}}} 
    \right)^{\frac{2}{7-q}}, \qquad
\rho^2 = \sum_{\alpha = q+1}^{9} (x^{\alpha})^{2}
\end{equation}
the metric for the transverse space $(U, \Omega_{8-q})$ 
in eq.\ (\ref{dq-bg_1}) can be written as 
\begin{eqnarray}
\left( \frac{R}{U}\right)^{\frac{7-q}{2}} \left( 
\frac{dU^{2}}{f(U)} + U^{2} d\Omega_{8-q}^{2} \right)
\A = \A K(\rho) \left( d\rho^{2} + \rho^{2} d\Omega_{8-q}^{2} 
\right) \nonu
\A = \A K(\rho) \sum_{\alpha=q+1}^{9} (dx^{\alpha})^{2}, 
\end{eqnarray}
where 
\begin{equation}
K(\rho) = \frac{R^{\frac{7-q}{2}} U^{\frac{q-3}{2}}}{\rho^{2}}. 
\end{equation}
We divide the coordinates $x^{q+1}, \cdots, x^9$ into two parts 
and introduce spherical coordinates $(\lambda, \Omega_{p-s-1})$ for 
the $x^{q+1}, \cdots, x^{q+p-s}$ directions and $(r, \Omega_{8-q-p+s})$ 
for the $x^{q+p-s+1}, \cdots, x^9$ directions.
Then the D$q$ background becomes
\begin{eqnarray}
ds^2 \A = \A \left( \frac{U}{R}\right)^{\frac{7-q}{2}}
    \left( -dt^2 + \sum_{i=1}^{q-1} (dx^i)^2 
        + f(U) (dx^{q})^2\right) \nonu
    \A\A + K(\rho) \left( d\lambda^{2} + \lambda^{2} d\Omega_{p-s-1}^{2}
                        + dr^2 + r^2 d\Omega_{8-q-p+s}^{2} \right),
\label{dq-bg_2}
\end{eqnarray}
where $\rho^2 = \lambda^2 + r^2$. We will wrap the probe D$p$-brane 
around S$^{p-s-1}$ in the next subsection.

%
\subsection{D$p$-brane embeddings}

We study the dynamics of a D$p$-brane in the D$q$ background.
In the limit $N_{c} \gg 1$ the D$p$-brane is introduced 
into the D$q$ background as a probe, which does not affect 
the background geometry.
The dynamics of the probe D$p$-brane in the background 
(\ref{dq-bg_2}) is described by the Dirac-Born-Infeld (DBI) action 
\begin{equation}
S_{\text{D}p} = - T_{p} \int d^{p+1}x \, 
    e^{-\phi} \sqrt{-\det g_{MN} },
\label{dbiaction}
\end{equation}
where $g_{MN}$ $(M, N = 0, 1, \cdots, p)$ is the induced metric 
on the world-volume and $T_{p}$ is the tension of the D$p$-brane. 
For simplicity we have ignored the gauge field on the probe D$p$-brane. 
\par
We use a physical gauge for D$p$-brane world-volume reparametrizations 
and use the spacetime coordinates $x^{\mu}$ $(\mu = 0,1,\cdots,s)$, 
$\lambda$, $\Omega_{p-s-1}$ as the world-volume coordinates.
Then the configurations of the D$p$-brane are determined by 
$x^{i}$ $(i=s+1, \cdots ,q)$, $r$ and $\Omega_{8-q-p+s}$
as a function of those world-volume coordinates.
We make an ansatz 
\begin{equation}
x^{s+1}, \cdots, x^q = \text{constant}, \quad 
r = r(\lambda), \quad 
\theta^{a} = \text{constant}, 
\label{ansatz}
\end{equation}
where $\theta^{a}$ ($a = 1, 2, \cdots, 8-q-p+s$) are coordinates 
of S$^{8-q-p+s}$.
\par
With this ansatz, the induced metric on the D$p$-brane is
\begin{equation}
ds^{2} = \left( \frac{U}{R} \right)^{\frac{7-q}{2}} 
         \eta_{\mu\nu} dx^\mu dx^\nu  
         + K(\rho) \left[ \left( 1 + (r')^{2} \right) d\lambda^{2} 
            + \lambda^{2} d\Omega_{p-s-1}^{2} \right],
\label{inducedmetric}
\end{equation}
where $r' = \frac{dr}{d\lambda}$.
Then the DBI action of the D$p$-brane becomes
\begin{equation}
S_{\text{D}p} 
= - \tilde{T_{p}} V_{p-s-1} \int d^{s+1}x \int d\lambda \; \rho^{\alpha}
  \left( 1 + \frac{U_{KK}^{7-q}}{4 \rho^{7-q}} \right)^{\beta}
  \lambda^{p-s-1} \sqrt{1 + (r')^{2}},
\label{action_dbi_low}
\end{equation}
where $\tilde{T_{p}} \equiv g_{s}^{-1} T_{p} R^{-\alpha}$ and
$V_{p-s-1}$ is the volume of S$^{p-s-1}$.
The parameters $\alpha$ and $\beta$ are defined as
\begin{equation}
\alpha = \frac{1}{4} (7-q) (4+2s-q-p) \;, \quad
\beta = \frac{1}{2} (4+2s-q-p) + \frac{2(p-s)}{7-q}.
\label{alphabeta}
\end{equation}
The action (\ref{action_dbi_low}) leads to the equation of 
motion for $r(\lambda)$
\begin{equation}
\frac{d}{d\lambda} \left[ \rho^{\alpha} 
        \left( 1 + \frac{U_{KK}^{7-q}}{4 \rho^{7-q}} \right)^{\beta}
            \frac{\lambda^{p-s-1} \; r'}{\sqrt{1 + (r')^{2}}}
\right] 
= \frac{\partial}{\partial r} \left[ \rho^{\alpha} 
        \left( 1 + \frac{U_{KK}^{7-q}}{4 \rho^{7-q}} \right)^{\beta} \right]
            \lambda^{p-s-1} \sqrt{1 + (r')^{2}}.
\label{eom_r_low_1}
\end{equation}
\par
As in refs.\ \cite{Babington:2003vm,Kruczenski:2003uq} we are 
interested in the situation in which the asymptotic distance between 
the D$q$-branes and the D$p$-brane is a finite constant $r_\infty$. 
This constant is proportional to the quark mass. 
Therefore we impose the boundary conditions for 
$\lambda \rightarrow \infty$
\begin{equation}
r(\lambda)|_{\lambda \rightarrow \infty} = r_{\infty}, \quad
r'(\lambda)|_{\lambda \rightarrow \infty} = 0.
\label{asymptoticbc}
\end{equation}
Then, eq.(\ref{eom_r_low_1}) can be linearized at large $\lambda$ as 
\begin{equation}
\frac{d}{d\lambda} \left( \lambda^{\alpha + p-s-1} r' \right)
    = \alpha \; \lambda^{\alpha + p-s-3} \, r,
\end{equation}
and the asymptotic behavior of the solution is
\begin{equation}
r(\lambda) \sim a \lambda^{k_{+}} + b \lambda^{k_{-}},
\label{r_asympt}
\end{equation}
where $a$, $b$ are constants and
\begin{equation}
k_{\pm} = \frac{-(\alpha +p-s-2) 
    \pm \sqrt{(\alpha + p-s-2)^{2} + 4 \alpha}}{2}.
\end{equation}
For the boundary condition (\ref{asymptoticbc}) to be satisfied, 
we must require $\alpha = 0$ and $p-s-2 > 0$. 
The first condition implies that the ground states of the NS 
sector of $q$-$p$ strings are massless since 
$\alpha = -2(7-q) a^{\text{NS}}$ as seen from eq.\ (\ref{zeropoint}). 
Then, the asymptotic behavior of $r(\lambda)$ is 
\begin{equation}
r(\lambda) \sim r_{\infty} + c \, \lambda^{-(p-s-2)}, 
\label{asymptoticr}
\end{equation}
where $c$ is a constant. 
As in ref.\ \cite{Kruczenski:2003uq} the quark condensate 
$\VEV{\bar{\psi}\psi}$ can be calculated by differentiating 
the vacuum energy density derived from the DBI action 
(\ref{action_dbi_low}) with respect to the quark mass $m_q$. 
Thus we obtain the quark mass and the quark condensate 
in terms of the constants $r_\infty$ and $c$ as 
\begin{equation}
m_{q} = \frac{r_{\infty}}{2\pi \ell_{s}^{2}}, \qquad
\VEV{\bar{\psi} \psi} 
= - 2 \pi (p-s-2) \, \ell_{s}^{2} \, \tilde{T}_{p} V_{p-s-1} \, c. 
\label{mass_cond}
\end{equation}
\par
\begin{figure}[!t]
\begin{center}
\includegraphics[width=7cm]{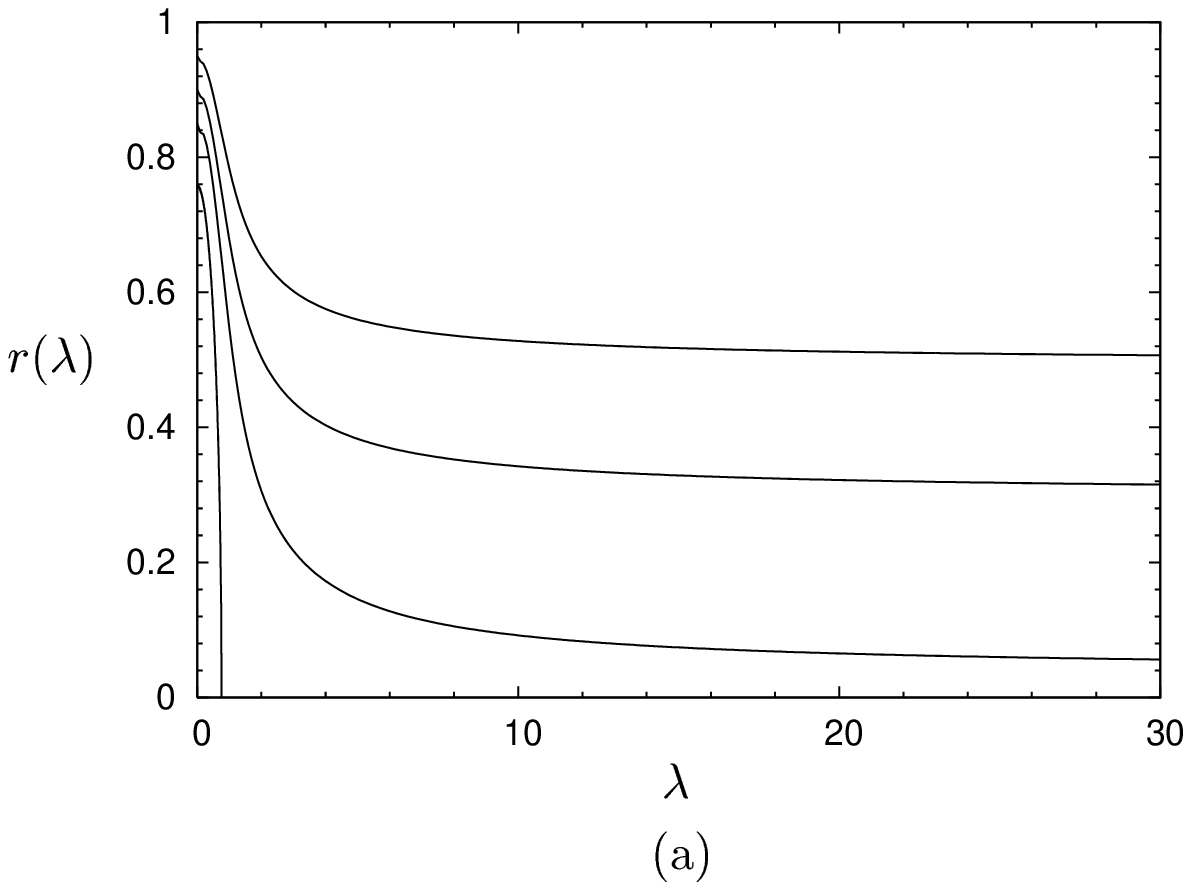} \hspace{3mm}
\includegraphics[width=7cm]{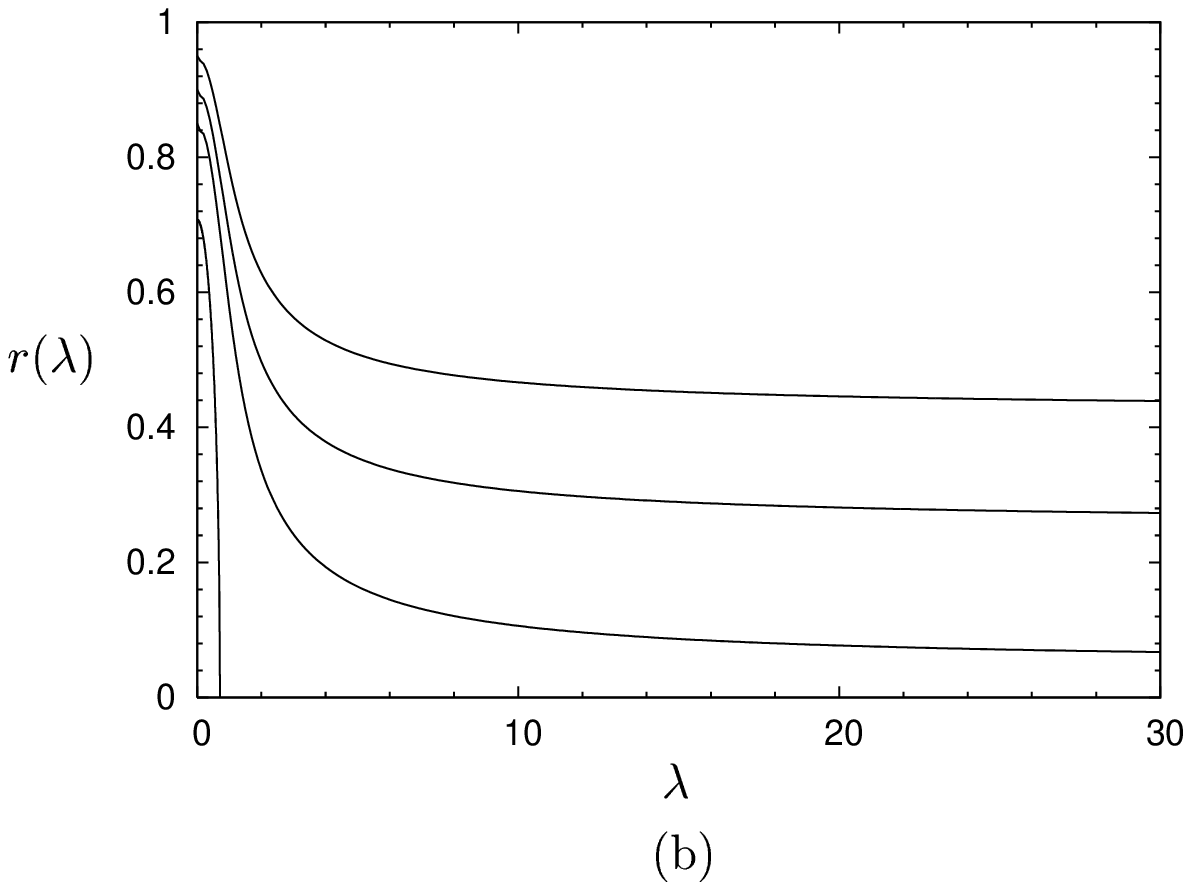} 
\caption[]{Solutions of eq.\ (\ref{eom_r_low_1}) 
for various values of $r_{\infty}$ in (a) the D2/D4 model 
with $s=1$ and (b) the D3/D5 model with $s=2$.} 
\label{solution_r}
\end{center}
\end{figure}

We have numerically solved eq.\ (\ref{eom_r_low_1}) for all 
possible values of $q$, $p$, $s$ satisfying $\alpha=0$, 
$p-s-2 > 0$, $s \leq 3$. 
The solutions of the D2/D4 model with $s=1$ and the 
D3/D5 model with $s=2$ are plotted in Fig.\ \ref{solution_r} 
for various values of $r_{\infty}$. 
The variables $\lambda$ and $r$ in these figures denote 
dimensionless ones rescaled by appropriate powers of $U_{KK}$. 
The leftmost curve in these figures represents $U = U_{KK}$. 
Its interior $U < U_{KK}$ is not a part of the space that we are 
considering. 
All the solutions have similar behaviors to those of the D4/D6 model 
with $s=3$, which was studied in ref.\ \cite{Kruczenski:2003uq}. 
The solutions approach a constant value $r_\infty$ for 
$\lambda = \infty$, while they reach a point outside of the 
curve $U = U_{KK}$ at $\lambda = 0$. 
The solutions break the rotational symmetry SO($9-q-p+s$) in the 
($r$, $\Omega_{8-q-p+s}$) space to SO($8-q-p+s$). 
\par
\begin{figure}[!t]
\begin{center}
\includegraphics[width=6.8cm]{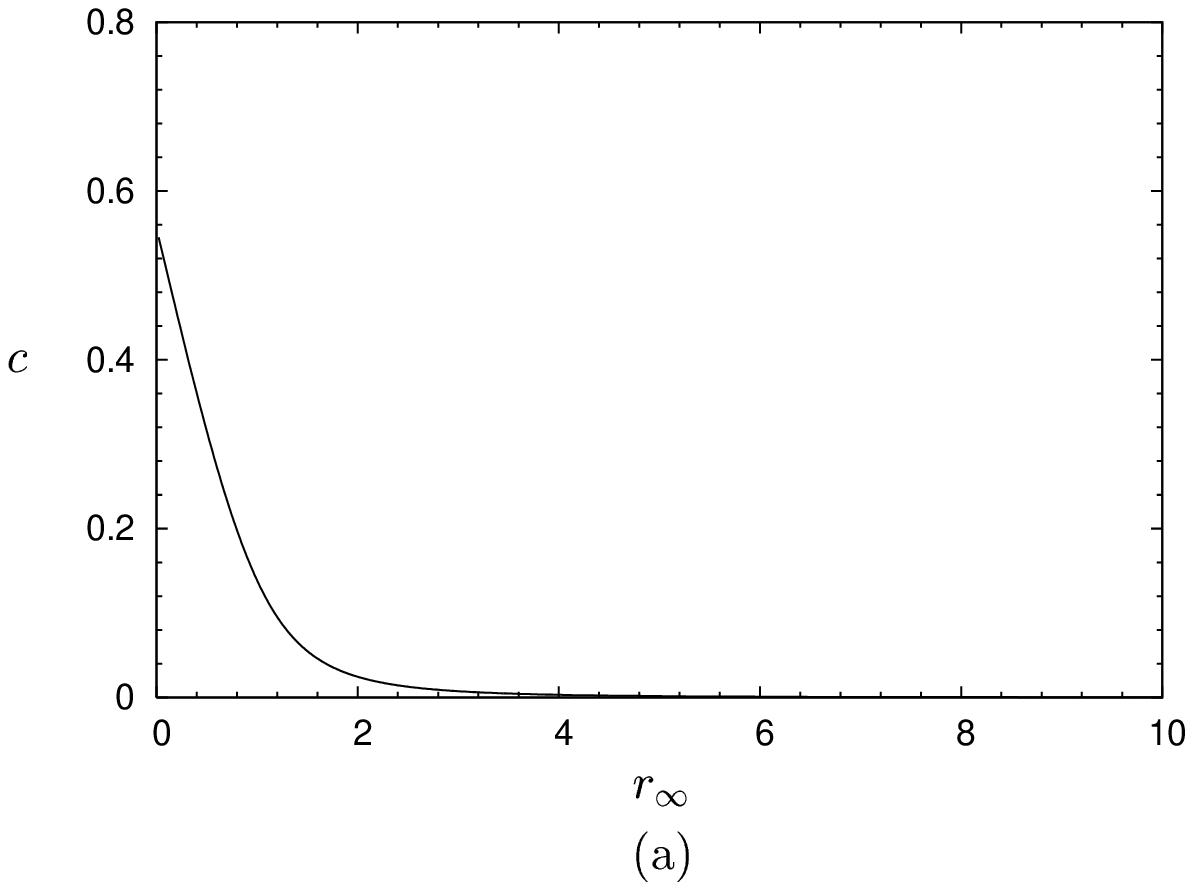} \hspace{4mm}
\includegraphics[width=6.8cm]{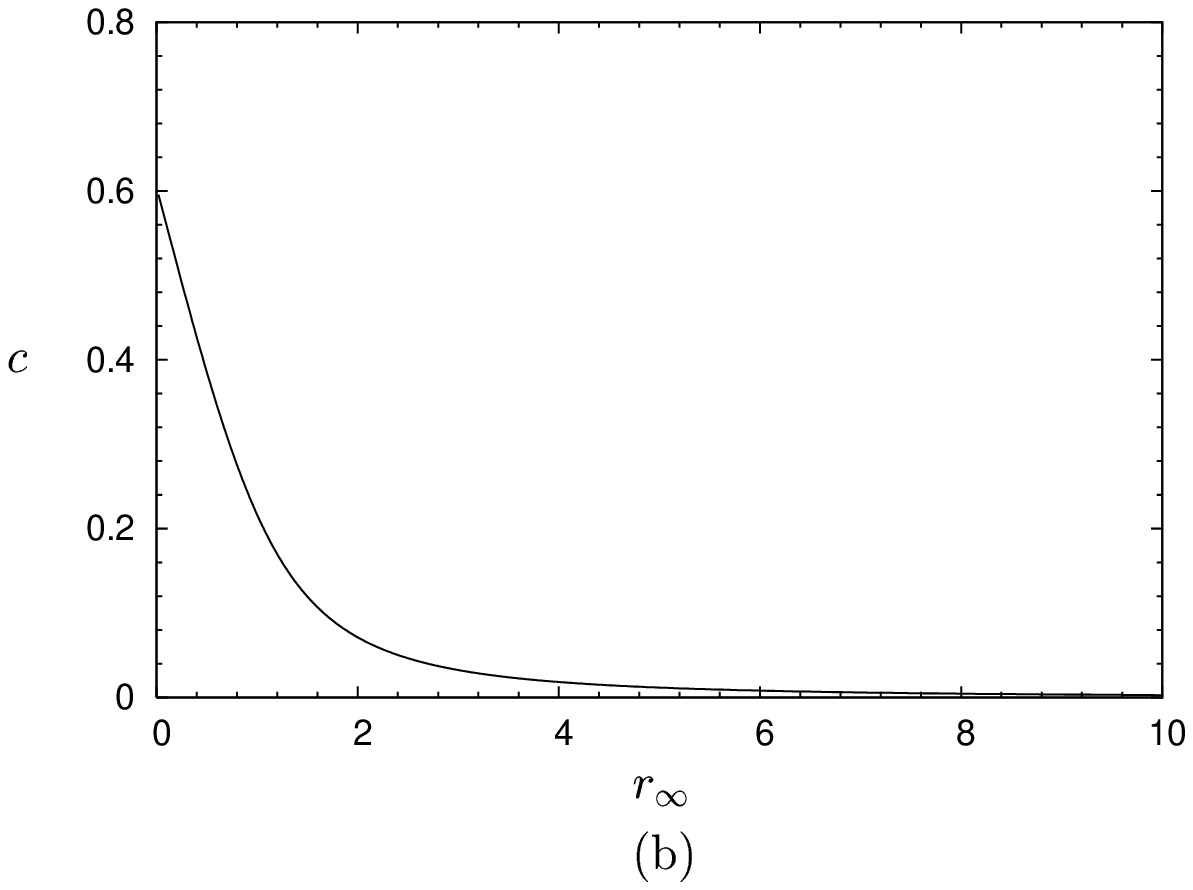} 
\caption[]{The quark condensate as a function of the quark mass 
for (a) the D2/D4 model with $s=1$ and 
(b) the D3/D5 model with $s=2$.}
\label{condensate}
\end{center}
\end{figure}

We have also numerically calculated the quark condensate as a 
function of the quark mass $c = c(r_{\infty})$ for all possible 
values of $q$, $p$, $s$ satisfying $\alpha=0$, $p-s-2 > 0$, 
$s \leq 3$. 
It is plotted in Fig.\ \ref{condensate} for the D2/D4 model 
with $s=1$ and the D3/D5 model with $s=2$. 
The variables $r_\infty$ and $c$ in these figures denote 
dimensionless ones rescaled by appropriate powers of $U_{KK}$. 
For all cases we find a non-zero quark condensate for $r_\infty = 0$. 
This agrees with a field theoretical view point. 
In QCD we expect that the chiral symmetry is spontaneously 
broken by the non-zero quark condensate even for $m_{q}=0$. 
\par
Finally, we write down a pattern of the symmetry breaking 
explicitly. 
%
For the D2/D4 model with $s=1$ it is 
\begin{equation}
\text{SU}(2)_{L} \times \text{SU}(2)_{R} \times \text{U}(1)_V
    \rightarrow \text{SU(2)}_{V} \times \text{U}(1)_V, 
\end{equation}
%
for the D3/D5 model with $s=2$ 
\begin{equation}
\text{SU}(2) \times \text{U}(1) 
\rightarrow \text{U}(1) \times \text{U}(1),
\end{equation}
and for the D4/D6 model with $s=3$ \cite{Kruczenski:2003uq} 
\begin{equation}
\text{U}(1)_{A} \times \text{U}(1)_{V} 
    \rightarrow \text{U(1)}_{V}. 
\end{equation}

%
\newsection{NG bosons as fluctuations of the probe brane}
In this section we study fluctuations of the probe brane 
around the vacuum embedding.
In the previous section we have seen that the vacuum embedding breaks 
SO($9-q-p+s$) rotational symmetry in the ($r, \Omega_{8-q-p+s}$) 
space to SO($8-q-p+s$).  This symmetry breaking corresponds 
to the chiral symmetry breaking for certain sets of $(q,p,s)$.
Therefore, there should be ($8-q-p+s$) Nambu-Goldstone (NG) bosons
associated with the symmetry breaking. 
If quarks are massless these bosons are massless NG bosons.
On the other hand, if quarks are massive these are pseudo-NG bosons 
with a non-vanishing mass. 
We will show that these pseudo-NG bosons satisfy the 
Gell-Mann-Oakes-Renner (GMOR) relation for a small quark mass 
$m_{q}$. We will also give the effective action of the 
fluctuations at quartic order. These results are a generalization 
of those of the D4/D6 system studied 
in ref.\ \cite{Kruczenski:2003uq} to the D$q$/D$p$ systems. 

%
\subsection{Fluctuations around the vacuum embeddings}

We study fluctuation modes around the vacuum D$p$-brane embedding 
\begin{equation}
x^{s+1}, \cdots, x^q = \text{constant}, \quad
r = r_{\text{vac}}(\lambda), \quad
\theta^{a} = 0 + \delta \theta^{a}(x^{M}),
\end{equation}
where $r_{\text{vac}}$ is the vacuum embedding determined 
numerically in the previous section. For simplicity we concentrate 
on fluctuations of $\theta^{a}$. In general, these 
fluctuations depend on all of the world-volume coordinates 
$x^{M}$ of the D$p$-brane. 
We will see that the fluctuations $\delta \theta^{a}$ are identified 
with the (pseudo-)NG bosons for the breaking of the rotational 
symmetry of S$^{8-q-p+s}$ (a subspace of the 
($r$, $\Omega_{8-q-p+s}$) space). 
\par
The induced metric on the D$p$-brane world-volume is
\begin{eqnarray}
ds^{2} 
\A = \A \left( \frac{U}{R} \right)^{\frac{7-q}{2}} 
        \eta_{\mu\nu} dx^\mu dx^\nu 
+ K(\rho) \left[ 
        \left( 1 + (r'_{\text{vac}})^{2} \right) d\lambda^{2} 
        + \lambda^{2} d\Omega_{p-s-1}^{2} \right] \nonu
\A\A \qquad\qquad 
    + K(\rho) r_{\text{vac}}^{2} \gamma_{ab} 
            \partial_{M} \delta\theta^{a} \partial_{N} \delta\theta^{b} 
          dx^{M} dx^{N} ,
\label{inducedmetric_fluctuation}
\end{eqnarray}
where $\rho^{2} = \lambda^{2} + r_{\text{vac}}^{2}$ 
and $\gamma_{ab}$ is the metric of a unit S$^{8-q-p+s}$.
Then the DBI action of the D$p$-brane (\ref{dbiaction}) 
to quadratic order becomes 
%
\begin{equation}
S_{\text{D}p} = S_{\text{vac}} + S_{\delta \theta},
\end{equation}
where $S_{\text{vac}}$ is the action for the vacuum embedding, 
i.e., eq.\ (\ref{action_dbi_low}) for $r = r_{\text{vac}}$, 
and $S_{\delta \theta}$ is the action for 
the fluctuations $\delta \theta^{a}$. 
%
After some simple calculations we obtain the action for 
$\delta\theta^a$ 
\begin{eqnarray}
S_{\delta \theta}
\A = \A - \tilde{T_{p}} \int d^{p+1}x 
    \sqrt{\det \gamma_{\alpha \beta}} \; 
    \lambda^{p-s-1} \sqrt{1 + (r'_{\text{vac}})^{2}} \nonu
\A\A \qquad\qquad 
    \times
     \left( 1 + \frac{U_{KK}^{7-q}}
     {4 \rho_{\text{vac}}^{7-q}} \right)^{\beta}
     \frac{K}{2} g^{MN} r_{\text{vac}}^{2} \gamma_{ab} 
     \partial_{M} \delta\theta^{a} \partial_{N} \delta\theta^{b} 
\label{action_theta}
\end{eqnarray}
and the equation of motion 
\begin{eqnarray}
\A\A \left( \frac{7-q}{2} \right)^{2} 
    \frac{U_{KK}^{5-q}}{M_{KK}^{2}} \rho_{\text{vac}}^{-(7-q)}
    \left( 1 + \frac{U_{KK}^{7-q}}{4 \rho_{\text{vac}}^{7-q}} 
        \right)^{\beta - \frac{2(5-q)}{7-q}}
    r_{\text{vac}}^{2}
    \partial_{\mu} \partial^{\mu} \delta\theta^{a} \nonu
\A\A \qquad\qquad 
+ \frac{1}{\lambda^{p-s-1} \sqrt{1 + (r'_{\text{vac}})^{2}}}
    \frac{\partial}{\partial\lambda}
    \left[
        \left( 1 + \frac{U_{KK}^{7-q}}
                    {4 \rho_{\text{vac}}^{7-q}} \right)^{\beta}
        \frac{\lambda^{p-s-1} r_{\text{vac}}^{2}} 
                {\sqrt{ 1 + (r'_{\text{vac}})^{2}}}
\frac{\partial}{\partial\lambda} \delta\theta^{a} 
    \right] \nonu
\A\A \qquad\qquad 
 + \left( 1 + \frac{U_{KK}^{7-q}}{4 \rho_{\text{vac}}^{7-q}} 
\right)^{\beta} \frac{r_{\text{vac}}^{2}}{\lambda^{2}}
\nabla^{2} \delta\theta^{a} 
= 0,
\label{eom_dtheta}
\end{eqnarray}
where $\gamma_{\alpha\beta}$ and $\nabla^2$ are the metric and 
the Laplacian on a unit S$^{p-s-1}$. 
\par
We can write a solution of the equation of motion 
(\ref{eom_dtheta}) in a form 
\begin{equation}
\delta \theta^{a} 
= F^{a}(\lambda) \, Y(\Omega_{p-s-1}) \, e^{ik \cdot x},
\label{theta}
\end{equation}
where $Y(\Omega_{p-s-1})$ is the spherical harmonics on S$^{p-s-1}$. 
We consider the zero (constant) mode of $Y$ and study only 
lowest-mass modes for simplicity. Substituting eq.\ (\ref{theta}) into 
eq.\ (\ref{eom_dtheta}) we obtain an eigenvalue equation for 
the $(s+1)$-dimensional mass $M^2_{\theta} = - k^{\mu}k_{\mu}$.
Although we can solve eq.\ (\ref{eom_dtheta}) by numerical 
calculations as in ref.\ \cite{Kruczenski:2003uq}, 
here we are content with asymptotic solutions of a linearized 
equation of motion for $\lambda \rightarrow \infty$. 
Taking account of the asymptotic behavior of $r_{\text{vac}}$ 
in eq.\ (\ref{asymptoticr}) the first term of eq.\ (\ref{eom_dtheta}) 
is sub-leading and the linearized equation 
for $\lambda \rightarrow \infty$ becomes 
\begin{equation}
\frac{\partial}{\partial\lambda} \left( \lambda^{p-s-1} r_{\text{vac}}^2 
\frac{\partial}{\partial\lambda} \delta\theta^a \right) = 0. 
\end{equation}
Depending on the value $r_\infty$ in eq.\ (\ref{asymptoticr}) 
the general solution is 
\begin{equation}
\delta\theta^a \sim \left\{
\begin{array}{@{\,}ll}
a \lambda^{p-s-2} + b & (r_\infty = 0) \\
a + b \lambda ^{-(p-s-2)} & (r_\infty \not= 0), 
\end{array}
\right.
\end{equation}
where $a$, $b$ are independent of $\lambda$. Since $p-s-2 > 0$, 
these solutions are normalizable when $a = 0$. 
\par
The non-linear equation of motion (\ref{eom_dtheta}) has exact 
solutions $\delta\theta^a = F^a e^{ik \cdot x}$ ($F^a =$ constant), 
which have an eigenvalue $M_\theta = 0$. From the above results 
on the asymptotic behaviors these solutions are normalizable 
only when $r_\infty = 0$. 
Since $r_{\infty} = 0$ means vanishing quark mass $m_{q}=0$, 
the normalizable solutions with $M_{\theta} = 0$ can be 
regarded as the NG bosons associated with the spontaneous breaking 
of the rotational symmetry $\text{SO}(9-q-p+s)$. 
When $r_\infty \not= 0$ ($m_q \not= 0$), the quark mass term 
explicitly breaks the chiral symmetry and we do not 
expect massless bosons. However, for small quark mass $m_q$ there 
should exist pseudo-NG bosons with a small mass $M_\theta$, 
which we consider in the next subsection. 
\par
In two dimensions there exists no massless NG boson associated 
with a spontaneous symmetry breaking \cite{Coleman:1973ci}. 
We have seen that there appear massless bosons even in the D2/D4 
model with a one-dimensional intersection when quarks are massless. 
These massless bosons should be an artifact of the large $N_c$ limit 
and should become massive if we take into account contributions 
from higher orders in the $1/N_{c}$ expansion. 
The situation is similar to the case of the Gross-Neveu 
model in two dimensions \cite{Gross:1974jv}, in which 
massless bosons appear in the large $N$ limit. 

%
\subsection{Light pseudo-NG bosons and the GMOR relation}

As we have seen above, the embeddings with $r_\infty = 0$ 
and those with $r_\infty \not= 0$ have different properties. 
For the $r_\infty = 0$ embeddings the asymptotic distance between 
D$q$ and D$p$-branes is zero and the quarks at the intersection are 
massless. There are ($8-q-p+s$) massless scalars $\delta\theta^{a}$ 
in the spectrum, which can be identified with the NG bosons 
associated with the spontaneous symmetry breaking 
SO($9-q-p+s$) $\rightarrow$ SO($8-q-p+s$).
We call these NG bosons \textit{pions}. 
On the other hand, for the $r_{\infty} \neq 0$ embeddings 
quarks are massive and the vacuum embedding explicitly breaks 
the rotational symmetry SO($9-q-p+s$) even for the asymptotic 
region $\lambda \rightarrow \infty$. 
In this case the fluctuations $\delta \theta^{a}$ are 
pseudo-NG bosons with a non-vanishing mass $M_\theta$. 
\par
We can show the Gell-Mann-Oakes-Renner (GMOR) relation 
\cite{Gell-Mann:1968rz} 
\begin{equation}
M_\theta^{2} = 
    - \frac{m_{q} \VEV{\bar{\psi}\psi}}{f_\pi^2}.
\label{gmor}
\end{equation}
for a small quark mass $m_q$ by using the holographic method 
\cite{Kruczenski:2003uq}. Here, $f_\pi$ is the pion decay constant. 
We begin with the $r_{\infty} =0$ embedding and make a small change 
$r_{\infty} = \delta r_{\infty}$. This gives a small mass to quarks.
As shown in ref.\ \cite{Kruczenski:2003uq} the mass of the 
pseudo-NG bosons $M_\theta$ can be obtained by using a standard 
perturbation theory in quantum mechanics and is written as 
\begin{equation}
M_\theta^{2} = (p-s-2) \, \frac{\bar{c} \, \delta r_{\infty}}
               {\int d\lambda \, \bar{\mu}}, 
\label{gmor2}
\end{equation}
where $\bar{c}$ is the coefficient in eq.\ (\ref{asymptoticr}) 
and $\bar{\mu}$ is given by 
\begin{equation}
\bar{\mu} =
    \left( \frac{7-q}{2} \right)^{2} \frac{U_{KK}^{5-q}}{M_{KK}^{2}}
    \bar{\rho}_{\text{vac}}^{-(7-q)} 
    \left( 1 + \frac{U_{KK}^{7-q}}{4 \bar{\rho}_{\text{vac}}^{7-q}} 
        \right)^{\beta - \frac{2(5-q)}{7-q}} 
    \bar{r}_{\text{vac}}^{2} \, \lambda^{p-s-1} 
        \sqrt{1 + (\bar{r}'_{\text{vac}})^{2}}.
\end{equation}
Here and in the following, quantities with a bar denote those 
for the $r_{\infty} = 0$ embedding. 
The quantities $\delta r_{\infty}$ and $\bar{c}$ are related to 
the quark mass and the quark condensate as in eq.\ (\ref{mass_cond}). 
The pion decay constant $f_\pi$ can be read from the effective action 
of $\delta \theta^{a}$. Assuming that $\delta \theta^{a}$ depend only 
on the coordinates of the intersection 
$x^\mu$ ($\mu = 0, 1, \cdots, s$) and integrating over $\lambda$ 
and $\Omega_{p-s-1}$ in eq.\ (\ref{action_theta}) we obtain 
\begin{equation}
S_{\delta \theta} = - f_\pi^2 \int d^{s+1}x \,
    \frac{1}{2} \gamma_{ab} \, \partial_{\mu} \delta\theta^{a} 
    \partial^{\mu} \delta\theta^{b}, 
\end{equation}
where $f_\pi$ is given by 
\begin{equation}
f_\pi^2 = \tilde{T}_{p} V_{p-s-1} 
    \int_{0}^{\infty} d \lambda \; \bar{\mu}. 
\label{decayconst}
\end{equation}
Using eqs.\ (\ref{mass_cond}), (\ref{decayconst}) in 
eq.\ (\ref{gmor2}), we obtain the GMOR relation (\ref{gmor}). 

%
\subsection{The pion effective action}

We can write down the effective action of the pion fields 
$\delta \theta^{a}$ at the intersection beyond the quadratic order. 
We assume that $\delta \theta^{a}$ depend 
only on the coordinates of the intersection $x^{\mu}$ 
($\mu = 0,1,\cdots,s$). 
By expanding the DBI action (\ref{dbiaction}) for the induced metric 
(\ref{inducedmetric_fluctuation}) to quartic order in $\delta\theta^a$ 
we obtain 
\begin{eqnarray}
S_{\delta \theta} 
\A = \A - \int d^{s+1}x \; 
\left( 
    \frac{f_\pi^2}{2} 
        \gamma_{ab} 
        \partial_{\mu}(\delta \theta^a) 
        \partial^{\mu}(\delta \theta^b) 
    + \frac{h_1}{4} 
        \left[ 
        \gamma_{ab} 
        \partial_{\mu}(\delta \theta^a) 
        \partial^{\mu}(\delta \theta^b) 
        \right]^{2} 
    \right. \nonu
\A\A \left.
    - \frac{h_2}{4} 
        \left[
        \gamma_{ab} 
        \partial_{\mu}(\delta \theta^a) 
        \partial_{\nu}(\delta \theta^b) 
        \right] 
        \left[
        \gamma_{cd}
        \partial^{\mu}(\delta \theta^c) 
        \partial^{\nu}(\delta \theta^d) 
        \right]
\right),
\end{eqnarray}
where $f_\pi$ is the pion decay constant (\ref{decayconst}) and 
the constants $h_1$, $h_2$ are given by 
\begin{eqnarray}
2 h_1 = h_2 
\A = \A \tilde{T_{p}} V_{p-s-1} 
    \left( \frac{7-q}{2} \right)^{4} \frac{U_{KK}^{2(5-q)}}{M_{KK}^{4}} 
    \int d\lambda \;
    \rho_{\text{vac}}^{-2(7-q)} \nonu
\A\A \times
    \left( 1 + \frac{U_{KK}^{7-q}}{4 \rho_{\text{vac}}^{7-q}} 
        \right)^{\beta - \frac{4(5-q)}{7-q}} 
    r_{\text{vac}}^{4} \, \lambda^{p-s-1} 
        \sqrt{1 + (r'_{\text{vac}})^{2}}. 
\end{eqnarray}
The relative coefficients of the quartic terms are different from 
those assumed in the Skyrme model 
\cite{Skyrme:1961vq,Zahed:1986qz} $h_1 = h_2$. 
This is in contrast with another approach \cite{Sakai:2004cn} 
to the holographic QCD, in which the relation $h_1 = h_2$ 
of the Skyrme model was obtained.

%
\newsection{Finite temperature analysis}
To study the theory at finite temperature we introduce a periodic 
Euclidean time coordinate $t_{E} \equiv it \sim t_{E} + \delta t_{E}$.
The period of $t_{E}$ is the inverse temperature $\delta t_{E} = 1/T$.
Then there are two periodic coordinates $t_{E}$ and $x^{q}$.
There exist two possible Euclidean geometries 
which have an appropriate asymptotic behavior.
One of them is the Euclidean version of eq.\ (\ref{dq-bg_1}). 
The other is the Euclidean version of the non-extremal black 
D$q$-brane geometry 
\begin{eqnarray}
ds^2 \A = \A \left( \frac{U}{R}\right)^{\frac{7-q}{2}} \!
    \left( \tilde{f}(U) dt_{E}^2 + \sum_{i=1}^{q-1} \left( dx^{i} \right)^2 
    + (dx^{q})^2\right) 
    + \left( \frac{R}{U}\right)^{\frac{7-q}{2}} \!
    \left( \frac{dU^2}{\tilde{f}(U)} + U^2 d\Omega_{8-q}^{2} \right), \nonu
\A\A \qquad\qquad\qquad \tilde{f}(U) = 1 - \left( \frac{U_{T}}{U} \right)^{7-q} 
\label{dq-bg_h1}
\end{eqnarray}
with the dilaton and the RR-flux given in eq.\ (\ref{dilaton_flux}). 
To avoid a conical singularity at $U=U_{T}$ in the $U$-$t_{E}$ plane 
the period $\delta t_{E}$ is fixed as
\begin{equation}
\delta t_{E} = \frac{4 \pi R^{\frac{7-q}{2}}}{ (7-q) \, 
U_{T}^{\frac{5-q}{2}}}.
\label{delta_t_e}
\end{equation}
It was shown \cite{Witten:1998zw,Kruczenski:2003uq,Aharony:2006da} 
that the Euclidean version of the background (\ref{dq-bg_1}) is 
dominant at low temperature, while the background (\ref{dq-bg_h1}) 
is dominant at high temperature by comparing values of the 
Euclidean supergravity action for these backgrounds.
A phase transition occurs at the temperature 
$T_{\text{deconf}} = M_{KK}/(2\pi)$. This phase transition corresponds 
to a confinement/deconfinement transition in the dual gauge theory 
\cite{Witten:1998zw}. 
\par
We consider the probe brane dynamics in the high temperature phase. 
The probe brane dynamics in the low temperature phase is essentially 
the same as at zero temperature. 
We only consider the models with $\alpha = 0$, $p-s-2 > 0$, $s \leq 3$ 
as in the zero temperature phase. 
With the ansatz (\ref{ansatz}) the induced metric on the D$p$-brane 
in the background (\ref{dq-bg_h1}) can be written as 
\begin{equation}
ds^{2} 
= \left( \frac{U}{R} \right)^{\frac{7-q}{2}} 
  \left( \tilde{f}(U) dt_{E}^{2} + \sum_{i=1}^{s}(dx^{i})^{2} \right) 
  + K(\rho) \left[ \left( 1 + (r')^{2} \right) d\lambda^{2} 
  + \lambda^{2} d\Omega_{p-s-1}^{2} \right].
\label{inducedmetric_h}
\end{equation}
%
Then the DBI action of the probe D$p$-brane becomes
\begin{equation}
S_{\text{D}p} 
= \tilde{T_{p}} V_{p-s-1} \int d^{s+1}x
  \int d\lambda 
  \left( 1 + \frac{U_{T}^{7-q}}{4 \rho^{7-q}} \right)^{\beta-1} 
  \left( 1 - \frac{U_{T}^{7-q}}{4 \rho^{7-q}} \right) 
  \lambda^{p-s-1} \sqrt{1 + (r')^{2}}, 
\label{action_dbi_h}
\end{equation}
which leads to the equation of motion for $r(\lambda)$
\begin{multline}
\frac{d}{d\lambda} \left[ 
\left( 1 + \frac{U_{T}^{7-q}}{4 \rho^{7-q}} \right)^{\beta-1}
\left( 1 - \frac{U_{T}^{7-q}}{4 \rho^{7-q}} \right)
\frac{\lambda^{p-s-1} \; r'}{\sqrt{1 + (r')^{2}}} \right] \\
= \frac{\partial}{\partial r} \left[ 
  \left( 1 + \frac{U_{T}^{7-q}}{4 \rho^{7-q}} \right)^{\beta-1}
  \left( 1 - \frac{U_{T}^{7-q}}{4 \rho^{7-q}} \right) \right]
  \lambda^{p-s-1} \sqrt{1 + (r')^{2}}.
\label{eom_r_h_2}
\end{multline}
%
\par
\begin{figure}[t]
\begin{center}
\includegraphics[width=7cm]{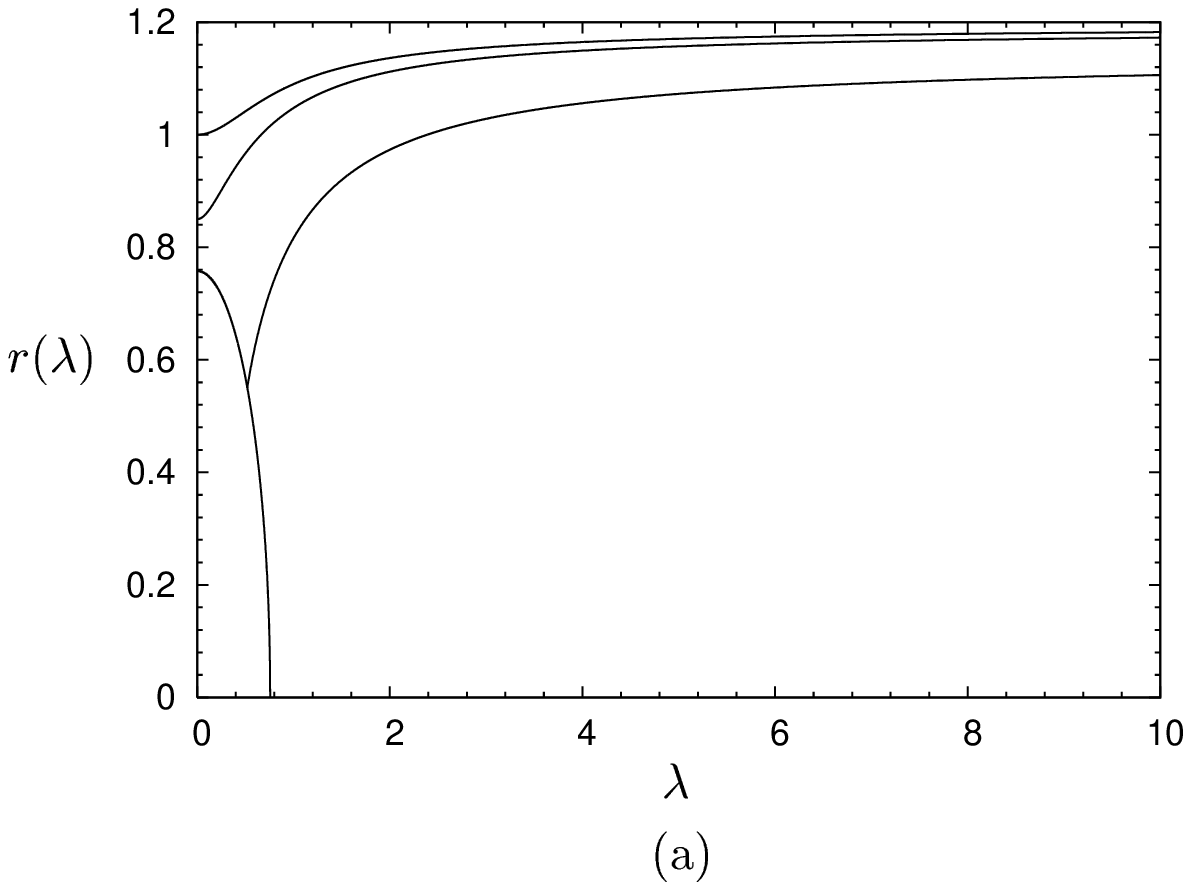} \hspace{3mm}
\includegraphics[width=7cm]{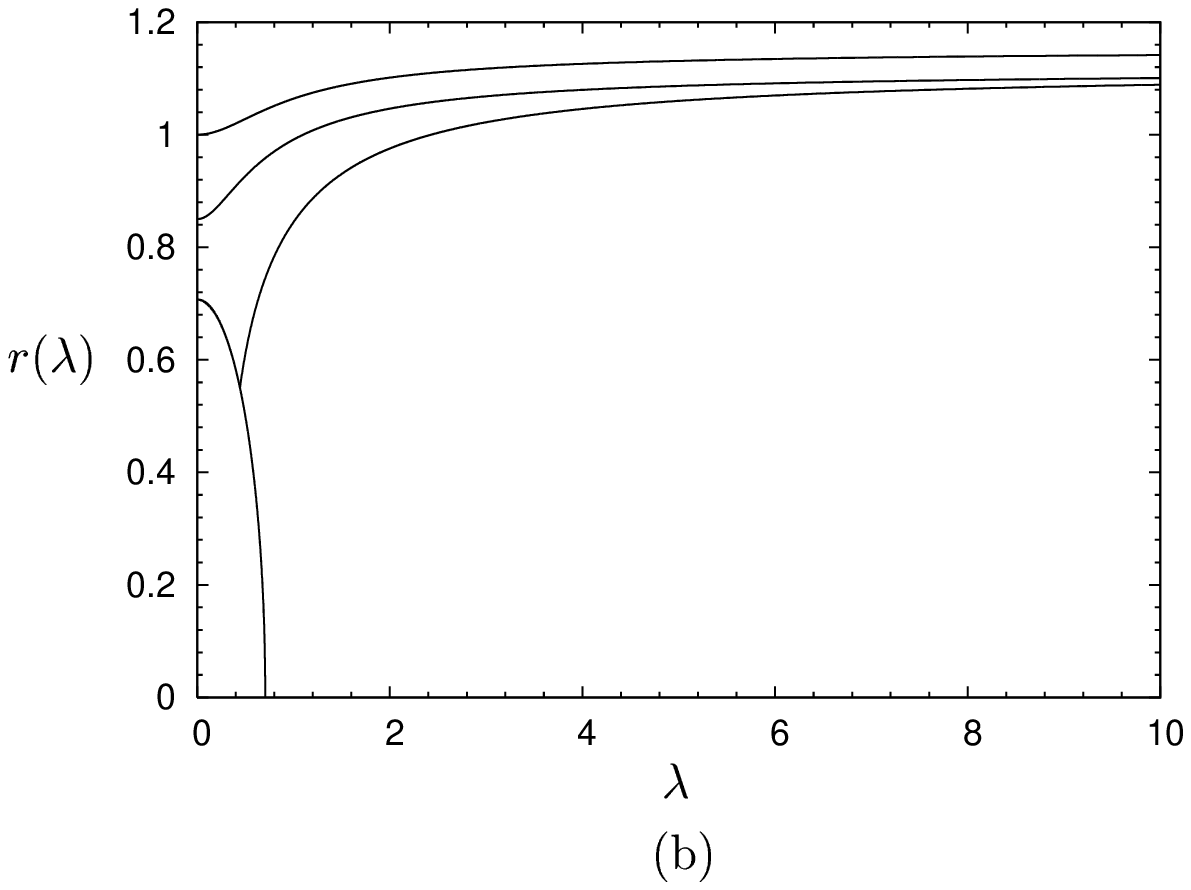} \\
\caption[]{Solutions of eq.\ (\ref{eom_r_h_2}) for various 
values of $r_{\infty}$ in (a) the D2/D4 model with $s=1$ and 
(b) the D3/D5 model with $s=2$.}
\label{solution_r_h}
\end{center}
\end{figure}

The asymptotic behavior of the solution $r(\lambda)$ of 
eq.\ (\ref{eom_r_h_2}) for large $\lambda$ is the same as 
in the zero temperature case (\ref{asymptoticr}). 
The parameters $r_{\infty}$ and $c$ are related to the quark 
mass $m_{q}$ and the quark condensate $\VEV{\bar{\psi}\psi}$ 
as in eq.\ (\ref{mass_cond}). 
%
We have numerically solved eq.\ (\ref{eom_r_h_2}) for all possible 
values of $q$, $p$, $s$. All the solutions have similar behaviors 
to those for the D4/D6 model with $s=3$ discussed in 
refs.\ \cite{Kruczenski:2003uq,Mateos:2006nu,Mateos:2007vn}. 
The solutions for the D2/D4 model with $s=1$ and the D3/D5 model 
with $s=2$ are plotted in Fig.\ \ref{solution_r_h} for various 
values of $r_{\infty}$. The variables $\lambda$ and $r$ in these 
figures (and Figs.\ \ref{condensate_h}, \ref{solution_r_h_s=q}, 
\ref{condensate_h_s=q} below) denote dimensionless 
ones rescaled by appropriate powers of $U_T$. 
The leftmost curve in these figures represents $U=U_{T}$.
\par
We have also numerically calculated the quark condensate 
$c = c(r_{\infty})$. 
Here we are interested in the phase structure of the system 
when the temperature $T$ is varied for fixed quark mass $m_{q}$.
The relation between $T$ and $r_{\infty}$ can be obtained 
from eqs.\ (\ref{delta_x_q}), (\ref{g_m_kk}), (\ref{delta_t_e}) as 
%
\begin{equation}
T = \frac{\bar{M}}{\sqrt{r_{\infty}^{5-q}}}, \qquad
\bar{M}^2 = \frac{(7-q)^2 m_q^{5-q} M_{KK}}{4(4\pi)^{\frac{5-q}{2}} 
\Gamma(\frac{7-q}{2}) g_q^2 N_c}, 
\end{equation}
where $g_{q}=g_{q+1}/\delta x^{q}$ is the $q$-dimensional 
Yang-Mills coupling and $r_\infty$ is the dimensionless 
variable rescaled by $U_T$. 
Using this relation the quark condensate as a function of the 
temperature is plotted in Fig.\ \ref{condensate_h}. 
\begin{figure}[t]
\begin{center}
\includegraphics[width=7cm]{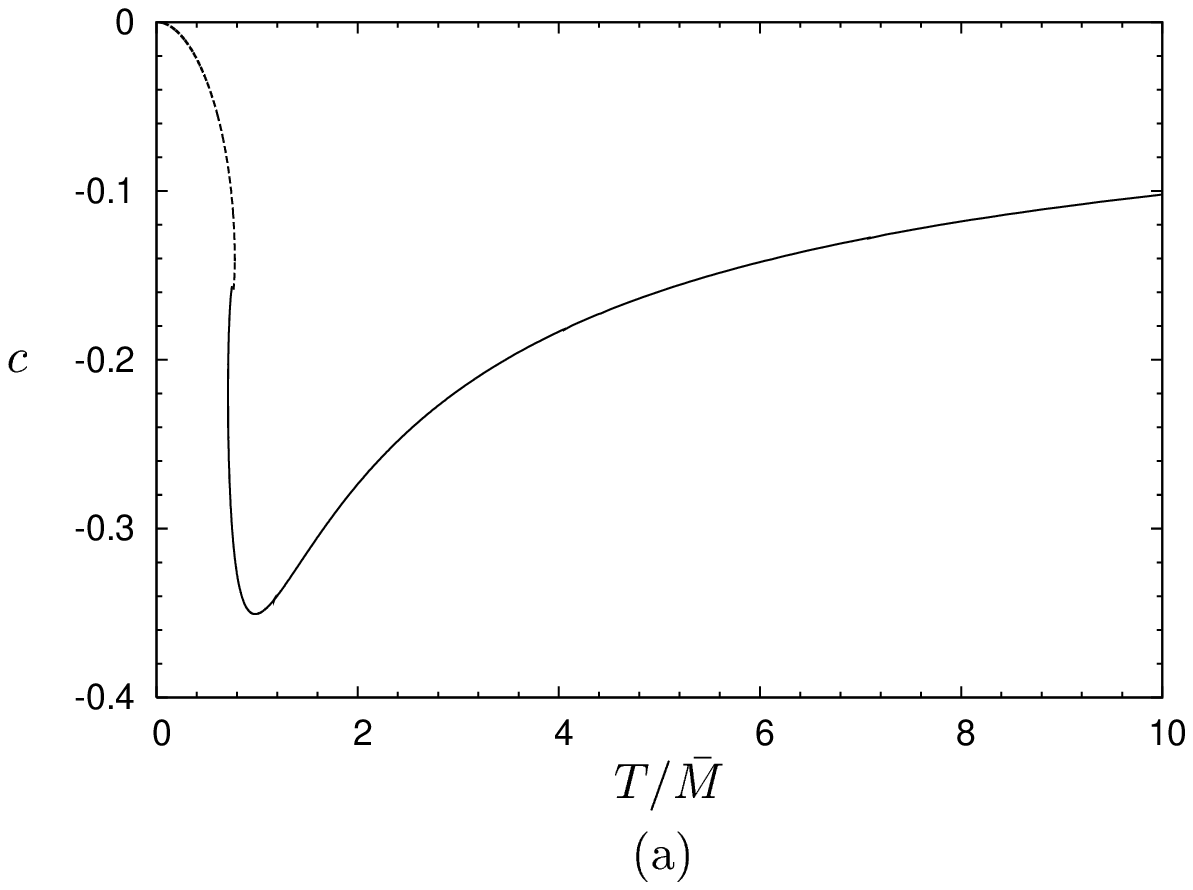} \hspace{3mm}
\includegraphics[width=7cm]{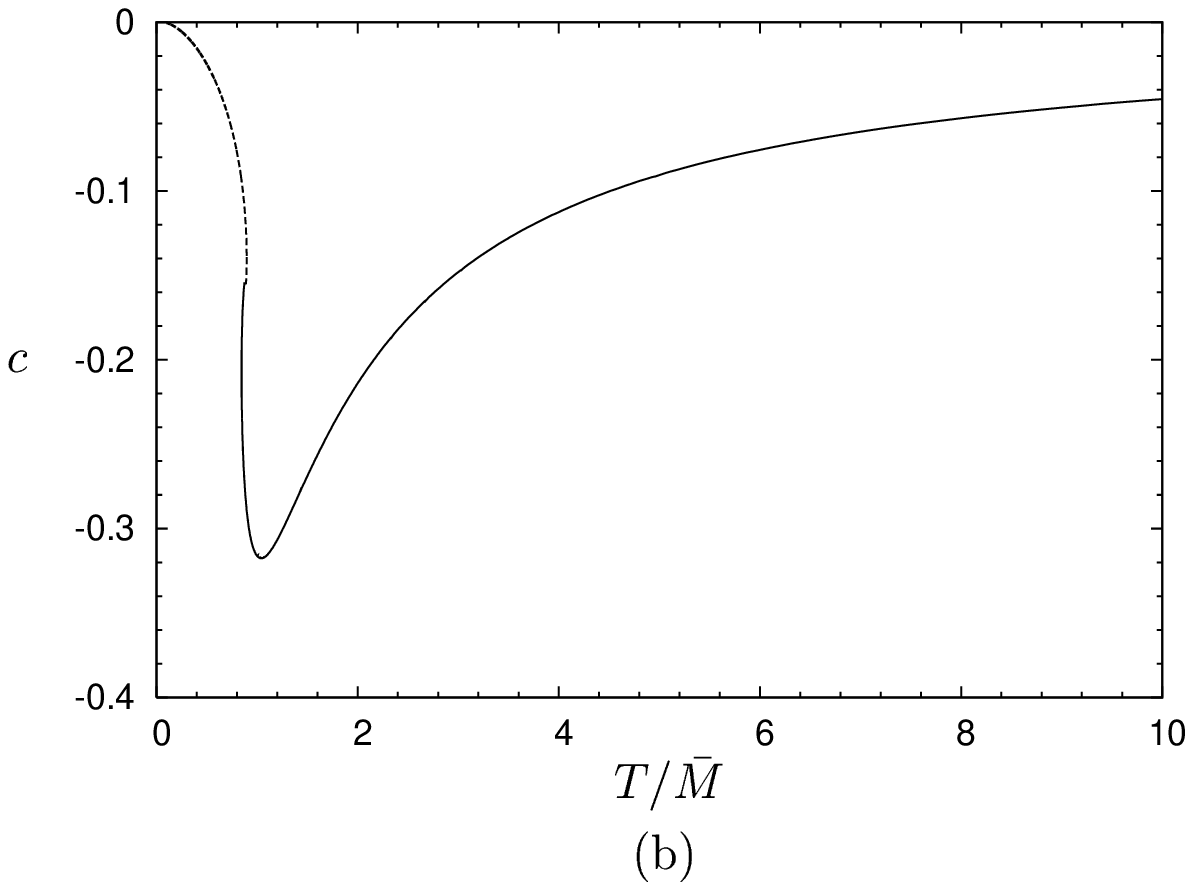} \\
\caption[]{The quark condensate as a function of temperature 
for (a) the D2/D4 model with $s=1$, (b) the D3/D5 model with $s=2$. 
The solid (dashed) lines represent the contributions 
from the embeddings which do (not) reach the horizon $U=U_T$.
}
\label{condensate_h}
\end{center}
\end{figure}
Note that the region near $T=0$ in these figures is not valid 
since the background (\ref{dq-bg_1}) is dominant at low temperature 
$T < T_{\text{deconf}}$. 
All the condensates have similar behaviors to those of the D4/D6 
model with $s=3$ discussed in 
refs.\ \cite{Kruczenski:2003uq,Mateos:2006nu,Mateos:2007vn}. 
\par
%
As was discussed in refs.\ \cite{Kruczenski:2003uq,Mateos:2006nu,
Mateos:2007vn} there are two types of embeddings. 
For sufficiently large $r_{\infty}$ the probe brane does not reach 
the horizon $U=U_T$. On the other hand, for sufficiently 
small $r_{\infty}$ it reaches the horizon $U=U_T$. 
For an intermediate region of $r_{\infty}$ more than one embeddings, 
which can be either type of embeddings, are possible. 
The physically realized embedding is the one with a minimal energy. 
Varying the value of $r_{\infty}$ a phase transition between these 
two types of embeddings occurs at a certain temperature 
$T=T_{\text{fund}}$. 
%
%
%
This phase transition, however, does not affect the chiral symmetry 
of the quarks in QCD$_{s+1}$ because of the non-zero value of $c$ 
for all temperature region except for $T \rightarrow \infty$. 
\par
%
%
The above finite temperature analysis can be applied to another type 
of brane configurations. 
Here we consider a non-compact limit $M_{KK} \rightarrow 0$ of the 
D$q$/D$(q+4)$ model with $s=q$ at finite temperature. 
At zero temperature it is dual to a supersymmetric gauge 
theory in $(q+1)$-dimensions. 
The case $q=3$ is the D3/D7 model at finite temperature discussed in 
refs.~\cite{Babington:2003vm,Mateos:2006nu,Mateos:2007vn}.
The D$q$/D$(q+4)$ configurations for $q=1$, 2, 3 are shown in 
Table~\ref{conf_qcd_q+1}.
\begin{table}[!t]
\begin{center}
\begin{tabular}{|c|c|cccccccccc|}
\hline
& & $0$ & $1$ & $2$ & $3$ & $4$ & 
        $5$ & $6$ & $7$ & $8$ & $9$  \\ \hline
color & D$1$ & $\circ$ & $\circ$ & $-$ & $-$ & $-$ & 
        $-$ & $-$ & $-$ & $-$ & $-$  \\
probe & D$5$ & $\circ$ & $\circ$ & $\circ$ & $\circ$ & $\circ$ & 
        $\circ$ & $-$ & $-$ & $-$ & $-$ \\
\hline
color & D$2$ & $\circ$ & $\circ$ & $\circ$ & $-$ & $-$ & 
        $-$ & $-$ & $-$ & $-$ & $-$ \\
probe & D$6$ & $\circ$ & $\circ$ & $\circ$ & $\circ$ & $\circ$ & 
        $\circ$ & $\circ$ & $-$ & $-$ & $-$  \\
\hline
color & D$3$ & $\circ$ & $\circ$ & $\circ$ & $\circ$ & $-$ & 
        $-$ & $-$ & $-$ & $-$ & $-$ \\
probe & D$7$ & $\circ$ & $\circ$ & $\circ$ & $\circ$ & $\circ$ & 
        $\circ$ & $\circ$ & $\circ$ & $-$ & $-$ \\
\hline
\end{tabular}
\caption[]{The D$q$/D$(q+4)$ brane configurations with $\text{\#ND}=4$.
}
\label{conf_qcd_q+1}
\end{center}
\end{table}
The rotational symmetry $\text{SO}(9-p)$ in the directions transverse 
to both branes is interpreted as a chiral symmetry for certain cases.
In particular, the $\text{SO}(4)_{6789}$ rotational symmetry 
of the D1/D5 model with $s=1$ is regarded as 
an $\text{SU}(2)_{L} \times \text{SU}(2)_{R}$ chiral symmetry 
and the $\text{SO}(2)_{89}$ rotational symmetry of the D3/D7 model 
with $s=3$ is regarded as an axial $\text{U}(1)_{A}$ symmetry 
\cite{Babington:2003vm}. 
\par
In the present case there are two possible background geometries.
One of them is the Euclidean 
$\text{AdS}_{q+2} \times \text{S}^{8-q}$, which is obtained by 
setting $U_{KK}=0$ in the Euclidean version of the 
metric (\ref{dq-bg_1}). 
The other is the Euclidean version of 
the Schwarzschild $\text{AdS}_{q+2} \times \text{S}^{8-q}$, which is 
given by the geometry (\ref{dq-bg_h1}) with non-compact $x^{q}$.
The phase transition occurs between these two backgrounds 
\cite{Witten:1998zw}. 
The Euclidean $\text{AdS}_{q+2} \times \text{S}^{8-q}$ is dominant 
at low temperature, while the Euclidean Schwarzschild 
$\text{AdS}_{q+2} \times \text{S}^{8-q}$ is dominant at high temperature.
\par
We consider the probe D$(q+4)$-brane dynamics in the high temperature 
phase. The induced metric and the equation of motion have the same 
form as (\ref{inducedmetric_h}) and (\ref{eom_r_h_2}) with $p=q+4$, $s=q$. 
The conditions $\alpha=0$, $p-s-2>0$, $s\le3$ require $q=1$, 2, 3. 
We have numerically solved eq.~(\ref{eom_r_h_2}) for these configurations.
All the solutions have similar behaviors to those for the D3/D7 model 
with $s=3$ \cite{Babington:2003vm,Mateos:2006nu,Mateos:2007vn}.
The solutions for the D1/D5 model with $s=1$ and the D2/D6 model with $s=2$ 
are plotted in Fig.~\ref{solution_r_h_s=q} for various values of $r_{\infty}$.
\begin{figure}[t]
\begin{center}
\includegraphics[width=7cm]{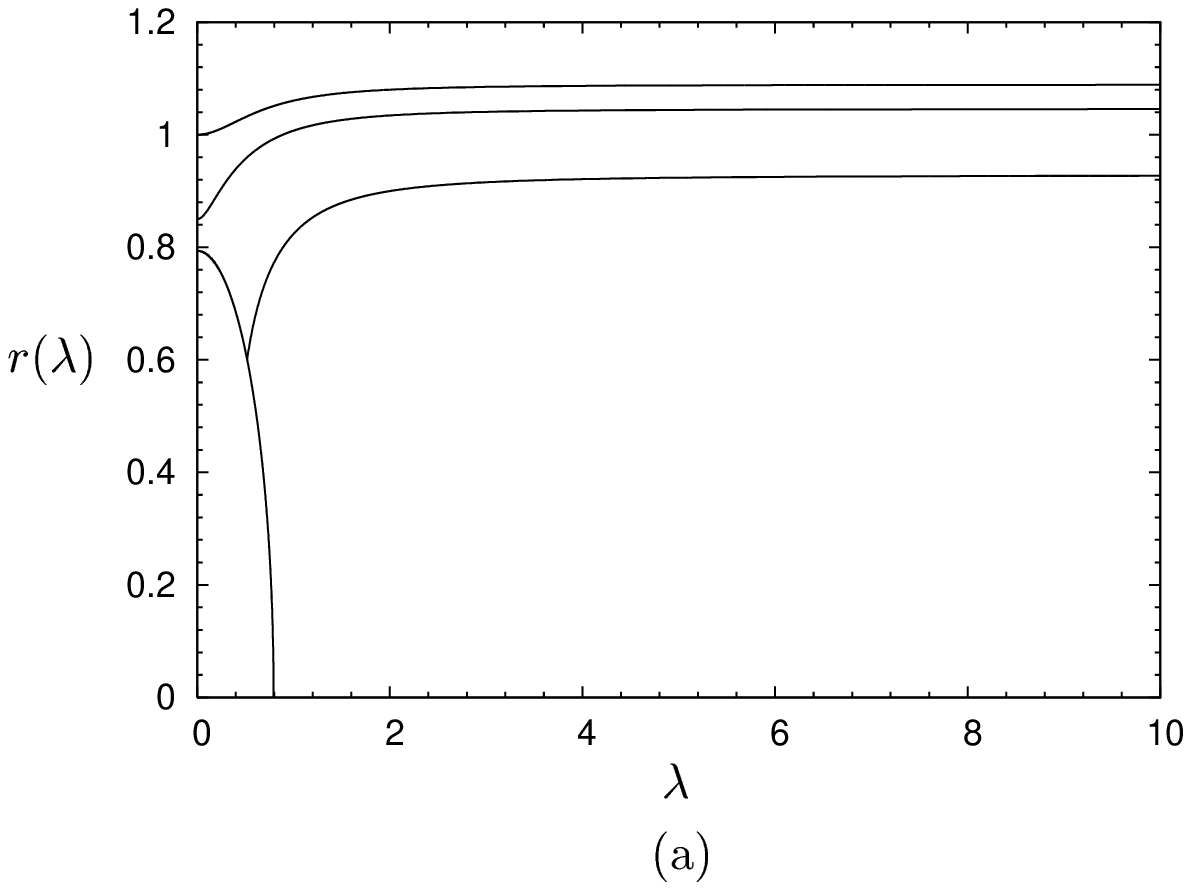} \hspace{3mm}
\includegraphics[width=7cm]{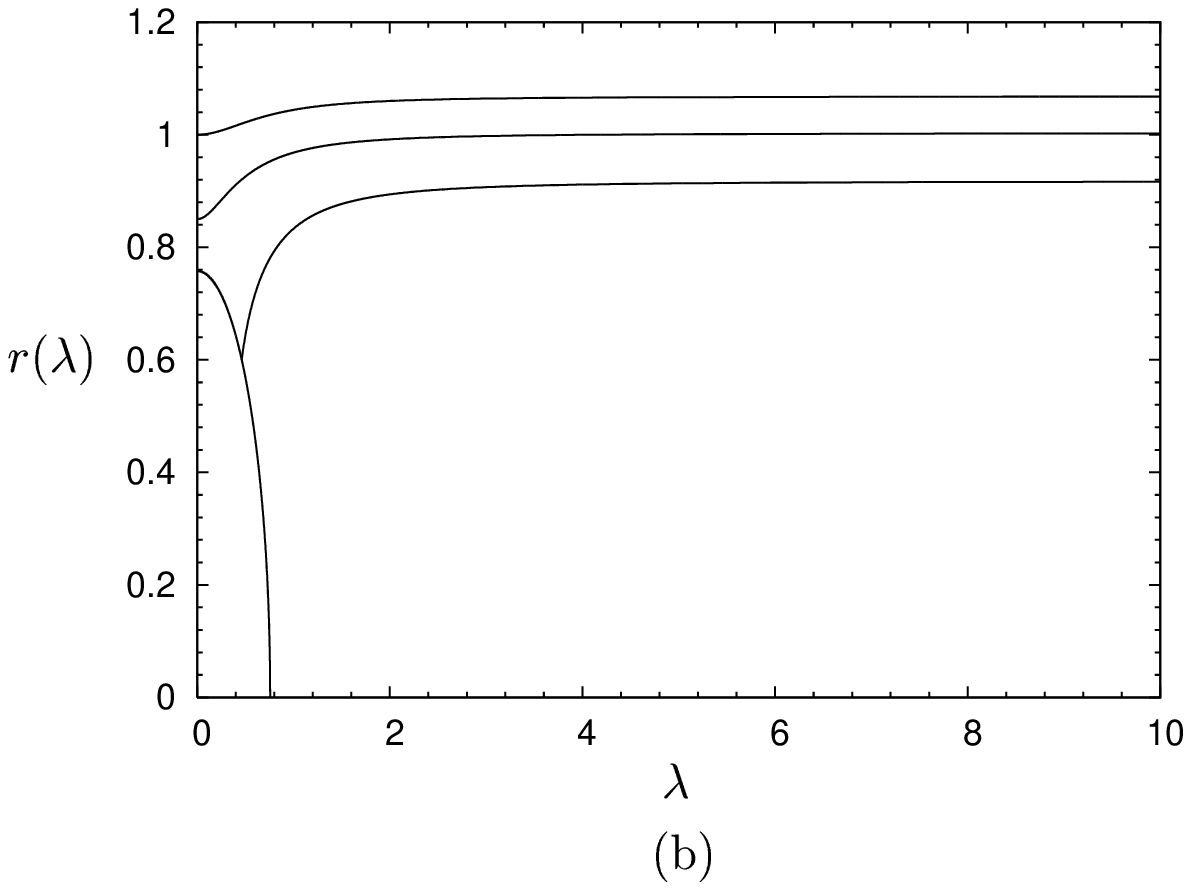} 
\caption[]{Solutions of eq.\ (\ref{eom_r_h_2}) for various 
values of $r_{\infty}$ in (a) the D1/D5 model with $s=1$ and 
(b) the D2/D6 model with $s=2$.}
\label{solution_r_h_s=q}
\end{center}
\end{figure}
We have also numerically calculated the quark condensate 
as a function of the temperature.
The results are plotted in Fig.~\ref{condensate_h_s=q}.
\begin{figure}[t]
\begin{center}
\includegraphics[width=7cm]{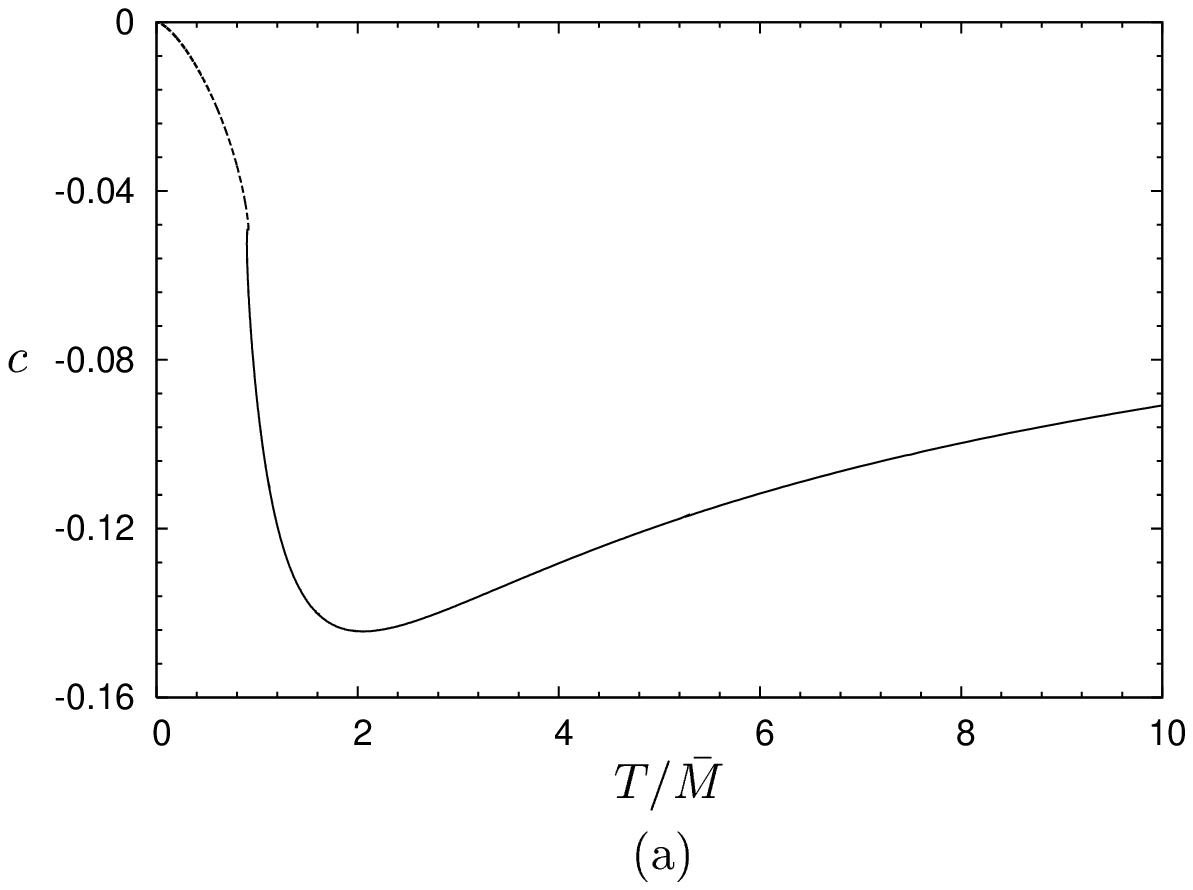} \hspace{3mm}
\includegraphics[width=7cm]{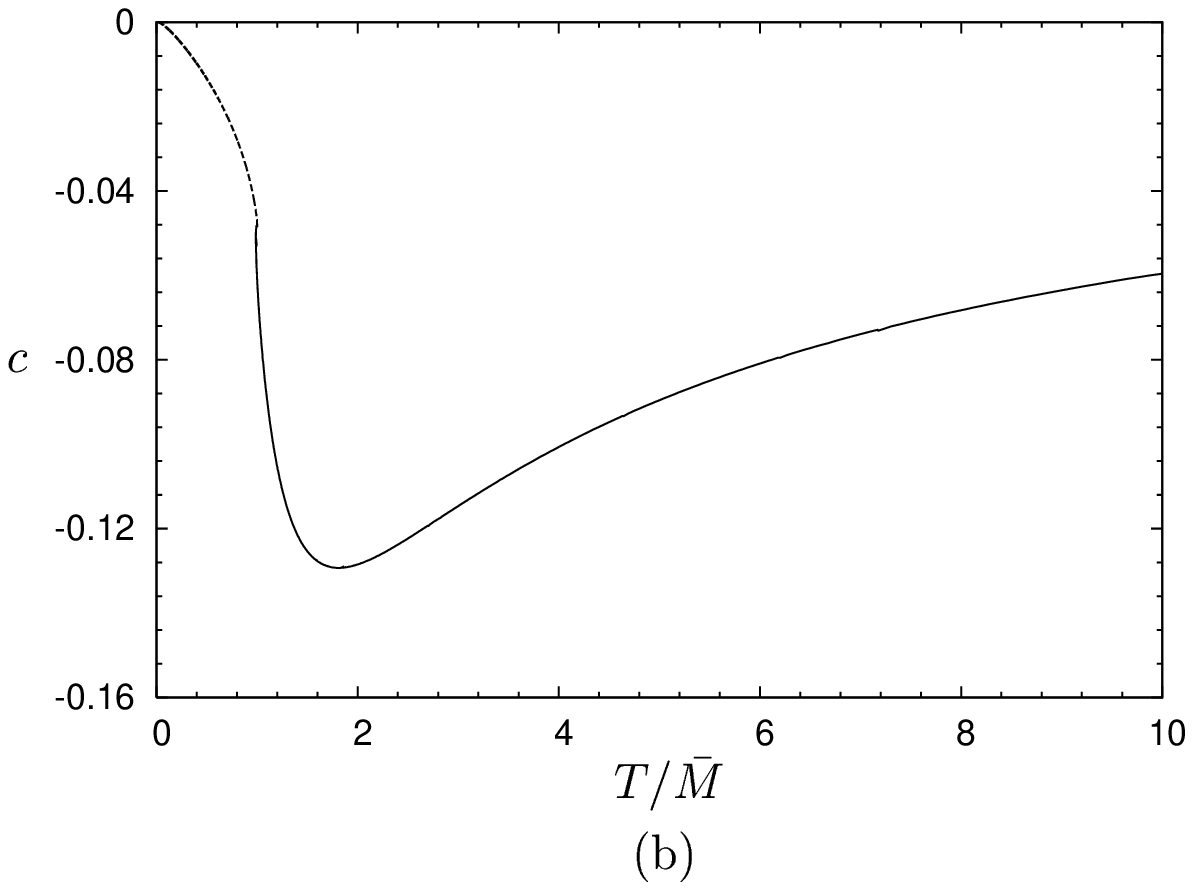} 
\caption[]{The quark condensate as a function of temperature 
for (a) the D1/D5 model with $s=1$, (b) the D2/D6 model with $s=2$.
The solid (dashed) lines represent the contributions 
from the embeddings which do (not) reach the horizon $U=U_T$.}
\label{condensate_h_s=q}
\end{center}
\end{figure}
All the condensates have similar behaviors to those of 
the D$3$/D$7$ model with $s=3$ discussed in 
refs.~\cite{Babington:2003vm,Mateos:2006nu,Mateos:2007vn}
\par
%
%
Finally we note that a chemical potential for the baryon number 
can be introduced by considering the U(1) gauge field on the probe 
D$p$-brane \cite{Kim:2006gp,Horigome:2006xu}. 
An asymptotically non-vanishing Euclidean time component 
of the U(1) gauge field can be understood as a chemical potential. 
It will be possible to discuss the phase diagram in the ($\mu$, $T$) 
space by using this chemical potential as in 
refs.\ \cite{Kobayashi:2006sb,Mateos:2007vc}. 

\bigskip 

%
\newsection{Conclusion}
In this paper we discussed the chiral symmetry breaking in 
the D$q$/D$p$ model with an $s$-dimensional intersection.
There exist QCD-like theories at the intersection for certain cases. 
We are interested in the models which have directions transverse 
to both of the D$q$ and D$p$-branes. 
The rotational symmetry in these directions can be identified 
with the chiral symmetry in certain cases. 
For instance, it is a non-Abelian chiral symmetry 
$\text{SU(2)}_{L} \times \text{SU(2)}_{R}$ 
in the D2/D4 model with $s=1$ corresponding to QCD$_{2}$. 
In these models we studied the dynamics of the probe D$p$-brane 
in the D$q$ background at zero and at finite temperature. 
\par
At zero temperature we found that the rotational symmetry in the 
transverse directions is broken by a D$p$-brane embedding. 
This corresponds to the chiral symmetry breaking in QCD$_{s+1}$. 
We numerically calculated the quark condensate 
$\VEV{\bar{\psi}\psi}$ as a function of the quark mass $m_{q}$. 
When $s < q $, we found that there is a non-zero quark condensate 
even for zero quark mass and therefore the symmetry is spontaneously 
broken. 
\par
We also studied the fluctuations around the vacuum embeddings. 
In the massless quark limit 
there appear $(8-q-p+s)$ massless scalar bosons, which are 
identified with the NG bosons associated with the 
spontaneous symmetry breaking. 
For massive quarks the symmetry is explicitly broken by a quark 
mass and there appear massive pseudo-NG bosons. 
We showed that the pseudo-NG bosons satisfy the GMOR relation 
for a small quark mass by using the holographic description. 
We also obtained the effective action of the NG bosons at quartic 
order. This action is different from the one assumed in the 
Skyrme model. 
\par
At finite temperature we found that the rotational symmetry is 
broken by the vacuum configuration as in the zero temperature case. 
This corresponds to the chiral symmetry breaking in QCD$_{s+1}$. 
We also found that there is a non-zero quark condensate 
$\VEV{\bar{\psi}\psi}$ except for $T \rightarrow \infty$. 
The quark condensate vanishes and the chiral symmetry is restored
only in the high temperature limit. 
It will be interesting to study the theory at finite chemical potential 
$\mu$ as well as at finite temperature $T$. Then we will be able to 
obtain the phase diagram in the $(\mu, T)$ space and discuss the 
chiral phase transition in the D$q$/D$p$ model. 

\bigskip

\vspace{10mm}
\noindent {\Large \textbf{Acknowledgements}} 
\vspace{3mm}

The work of N.H. is partly supported by the Sasakawa Scientific 
Research Grant from The Japan Science Society.

\bigskip

%


\begin{thebibliography}{100}
%
%
\bibitem{Maldacena:1997re}
    J.~Maldacena,
    The large $N$ limit of superconformal field theories and 
    supergravity, \textit{Adv.\ Theor.\ Math.\ Phys.\ }\textbf{2} 
    (1998) 231 [arXiv:hep-th/9711200].
%
\bibitem{Gubser:1998bc}
    S.S.~Gubser, I.R.~Klebanov and A.M.~Polyakov,
    Gauge theory correlators from non-critical string theory, 
    \textit{Phys.\ Lett.\ }\textbf{B428} (1998) 105 
    [arXiv:hep-th/9802109].
%
\bibitem{Witten:1998qj}
    E.~Witten, Anti de Sitter space and holography, 
    \textit{Adv.\ Theor.\ Math.\ Phys.\ }\textbf{2} (1998) 253 
    [arXiv:hep-th/9802150].
%
\bibitem{Aharony:1999ti}
    O.~Aharony, S.S.~Gubser, J.~Maldacena, H.~Ooguri and Y.~Oz,
    Large $N$ field theories, string theory and gravity, 
    \textit{Phys.\ Rept.\ }\textbf{323} (2000) 183 
    [arXiv:hep-th/9905111].
%
%
\bibitem{Karch:2002sh}
    A.~Karch and E.~Katz, Adding flavor to AdS/CFT,
    \textit{JHEP} \textbf{0206} (2002) 043 
    [arXiv:hep-th/0205236].
%
\bibitem{Sakai:2003wu}
    T.~Sakai and J.~Sonnenschein,
    Probing flavored mesons of confining gauge theories by supergravity,
    \textit{JHEP} \textbf{0309} (2003) 047 
    [arXiv:hep-th/0305049]
%
\bibitem{Babington:2003vm}
    J.~Babington, J.~Erdmenger, N.J.~Evans, Z.~Guralnik and I.~Kirsch,
    Chiral symmetry breaking and pions in non-supersymmetric 
    gauge/gravity duals,
    \textit{Phys.\ Rev.\ }\textbf{D69} (2004) 066007 
    [arXiv:hep-th/0306018].
%
\bibitem{Kruczenski:2003uq}
    M.~Kruczenski, D.~Mateos, R.C.~Myers and D.J.~Winters,
    Towards a holographic dual of large-$N_c$ QCD,
    \textit{JHEP} \textbf{0405} (2004) 041 
    [arXiv:hep-th/0311270].
%
\bibitem{Evans:2004ia}
    N.J.~Evans and J.P.~Shock, Chiral dynamics from AdS space,
    \textit{Phys.\ Rev.\ }\textbf{D70} (2004) 046002 
    [arXiv:hep-th/0403279].
%
\bibitem{Ghoroku:2004sp}
    K.~Ghoroku and M.~Yahiro,
    Chiral symmetry breaking driven by dilaton, 
    \textit{Phys.\ Lett.\ }\textbf{B604} (2004) 235 
    [arXiv:hep-th/0408040].
%
\bibitem{Bak:2004nt}
    D.~Bak and H.~U.~Yee, 
    Separation of spontaneous chiral symmetry breaking and confinement via 
    AdS/CFT correspondence,
    \textit{Phys.\ Rev.\ }\textbf{D71} (2005) 046003 
    [arXiv:hep-th/0412170].
%
\bibitem{Rodriguez:2005jr}
    M.J.~Rodriguez and P.~Talavera,
    A 1+1 field theory spectrum from M theory,
    arXiv:hep-th/0508058.
%
%
\bibitem{Sakai:2004cn}
    T.~Sakai and S.~Sugimoto, 
    Low energy hadron physics in holographic QCD, 
    \textit{Prog.\ Theor.\ Phys.\ }\textbf{113} (2005) 843 
    [arXiv:hep-th/0412141].
%
\bibitem{Sakai:2005yt}
    T.~Sakai and S.~Sugimoto,
    More on a holographic dual of QCD, 
    \textit{Prog.\ Theor.\ Phys.\ }\textbf{114} (2006) 1083 
    [arXiv:hep-th/0507073].
%
\bibitem{Antonyan:2006vw}
    E.~Antonyan, J.A.~Harvey, S.~Jensen and D.~Kutasov, 
    NJL and QCD from string theory, 
    arXiv:hep-th/0604017.
%
\bibitem{Gao:2006up}
    Y.h.~Gao, W.s.~Xu and D.f.~Zeng,
    NGN, QCD${}_2$ and chiral phase transition from string theory,
    \textit{JHEP} \textbf{0608} (2006) 018
    [arXiv:hep-th/0605138].
%
\bibitem{Antonyan:2006qy}
    E.~Antonyan, J.A.~Harvey and D.~Kutasov,
    The Gross-Neveu model from string theory,
    \textit{Nucl.\ Phys.\ }\textbf{B776} (2007) 93
    [arXiv:hep-th/0608149].
%
\bibitem{Basu:2006eb}
    A.~Basu and A.~Maharana,
    Generalized Gross-Neveu models and chiral symmetry breaking 
    from string theory,
    \textit{Phys.\ Rev.\ }\textbf{D75} (2007) 065005
    [arXiv:hep-th/0610087].
%
\bibitem{Antonyan:2006pg}
    E.~Antonyan, J.A.~Harvey and D.~Kutasov,
    Chiral symmetry breaking from intersecting D-branes,
    \textit{Nucl.\ Phys.\ }\textbf{B784} (2007) 1
    [arXiv:hep-th/0608177].
%
\bibitem{Gepner:2006qy}
    D.~Gepner and S.S.~Pal,
    Chiral symmetry breaking and restoration from holography,
    arXiv:hep-th/0608229.
%
%
\bibitem{Casero:2007ae}
    R.~Casero, E.~Kiritsis and A.~Paredes,
    Chiral symmetry breaking as open string tachyon condensation, 
    \textit{Nucl.\ Phys.\ }\textbf{B787} (2007) 98 [arXiv:hep-th/0702155].
%
\bibitem{Hashimoto:2007fa}
    K.~Hashimoto, T.~Hirayama and A.~Miwa,
    Holographic QCD and pion mass, \textit{JHEP} {\bf 0706} (2007) 020
    [arXiv:hep-th/0703024].
%
\bibitem{Evans:2007jr}
    N.~Evans and E.~Threlfall,
    Quark mass in the Sakai-Sugimoto model of chiral symmetry 
    breaking, arXiv:0706.3285 [hep-th].
\bibitem{Bergman:2007pm}
    O.~Bergman, S.~Seki and J.~Sonnenschein,
    Quark mass and condensate in HQCD, arXiv:0708.2839 [hep-th].
\bibitem{Dhar:2007bz}
    A.~Dhar and P.~Nag, 
    Sakai-Sugimoto model, tachyon condensation and chiral 
    symmetry breaking, arXiv:0708.3233 [hep-th].
%
%
\bibitem{Ghoroku:2005tf}
    K.~Ghoroku, T.~Sakaguchi, N.~Uekusa and M.~Yahiro,
    Flavor quark at high temperature from a holographic model, 
    \textit{Phys.\ Rev.\ }\textbf{D71} (2005) 106002
    [arXiv:hep-th/0502088].
%
\bibitem{Aharony:2006da} 
    O.~Aharony, J.~Sonnenschein and S.~Yankielowicz,
    A holographic model of deconfinement and chiral symmetry 
    restoration, \textit{Annals Phys.\ }\textbf{322} (2007) 1420
    [arXiv:hep-th/0604161].
%
\bibitem{Parnachev:2006dn} 
    A.~Parnachev and D.A.~Sahakyan,
    Chiral phase transition from string theory, 
    \textit{Phys.\ Rev.\ Lett.\ }\textbf{97} (2006) 111601 
    [arXiv:hep-th/0604173].
%
\bibitem{Mateos:2006nu}
    D.~Mateos, R.C.~Myers and R.M.~Thomson,
    Holographic phase transitions with fundamental matter, 
    \textit{Phys.\ Rev.\ Lett.\ }\textbf{97} (2006) 091601
    [arXiv:hep-th/0605046].
%
\bibitem{Peeters:2006iu} 
    K.~Peeters, J.~Sonnenschein and M.~Zamaklar,
    Holographic melting and related properties of mesons 
    in a quark gluon plasma,
    \textit{Phys.\ Rev.\ }\textbf{D74} (2006) 106008
    [arXiv:hep-th/0606195].
%
\bibitem{Mateos:2007vn}
    D.~Mateos, R.C.~Myers and R.M.~Thomson,
    Thermodynamics of the brane,
    \textit{JHEP} \textbf{0705} (2007) 067
    [arXiv:hep-th/0701132].
%
%
\bibitem{Kim:2006gp} 
    K.Y.~Kim, S.J.~Sin and I.~Zahed, 
    Dense hadronic matter in holographic QCD, 
    arXiv:hep-th/0608046.
%
\bibitem{Horigome:2006xu}
    N.~Horigome and Y.~Tanii,
    Holographic chiral phase transition with chemical potential,
    \textit{JHEP} \textbf{0701} (2007) 072
    [arXiv:hep-th/0608198].
%
\bibitem{Parnachev:2006ev} 
    A.~Parnachev and D.A.~Sahakyan,
    Photoemission with chemical potential from QCD gravity dual, 
    \textit{Nucl.\ Phys.\ }\textbf{B768} (2007) 177
    [arXiv:hep-th/0610247].
%
\bibitem{Nakamura:2006xk}
    S.~Nakamura, Y.~Seo, S.J.~Sin and K.P.~Yogendran,
    A new phase at finite quark density from AdS/CFT, 
    arXiv:hep-th/0611021.
%
\bibitem{Kobayashi:2006sb}
    S.~Kobayashi, D.~Mateos, S.~Matsuura, R.C.~Myers and R.M.~Thomson,
    Holographic phase transitions at finite baryon density,
    \textit{JHEP} \textbf{0702} (2007) 016
    [arXiv:hep-th/0611099].
%
\bibitem{Mateos:2007vc}
    D.~Mateos, S.~Matsuura, R.C.~Myers and R.M.~Thomson,
    Holographic phase transitions at finite chemical potential, 
    {\it JHEP} {\bf 0711} (2007) 085 [arXiv:0709.1225 [hep-th]].
%
%
\bibitem{DeWolfe:2001pq}
  O.~DeWolfe, D.Z.~Freedman and H.~Ooguri, 
  Holography and defect conformal field theories, 
  \textit{Phys.\ Rev.} \textbf{D66} (2002) 025009 
  [arXiv:hep-th/0111135].
%
\bibitem{Constable:2002xt}
  N.R.~Constable, J.~Erdmenger, Z.~Guralnik and I.~Kirsch, 
  Intersecting D3-branes and holography, 
  \textit{Phys.\ Rev.} \textbf{D68} (2003) 106007 
  [arXiv:hep-th/0211222].
%
%
\bibitem{Gell-Mann:1968rz}
    M.~Gell-Mann, R.J.~Oakes and B.~Renner,
    Behavior of current divergences under SU(3) $\times$ SU(3),
    \textit{Phys.\ Rev.\ }\textbf{175} (1968) 2195.
%
\bibitem{Polchinski:1998rr}
    J.~Polchinski,
    \textit{String Theory.\ Vol.\ 2: Superstring Theory and Beyond}
    (Cambridge Univ.\ Press, 1998)
%
\bibitem{Itzhaki:1998dd}
    N.~Itzhaki, J.M.~Maldacena, J.~Sonnenschein and S.~Yankielowicz,
    Supergravity and the large $N$ limit of theories with sixteen 
    supercharges, \textit{Phys.\ Rev.\ }\textbf{D58} (1998) 046004
    [arXiv:hep-th/9802042].
%
\bibitem{Coleman:1973ci}
    S.R.~Coleman, There are no Goldstone bosons in two-dimensions,
    \textit{Commun.\ Math.\ Phys.\ }\textbf{31} (1973) 259.
%
\bibitem{Gross:1974jv}
    D.J.~Gross and A.~Neveu, Dynamical symmetry breaking 
    in asymptotically free field theories, 
    \textit{Phys.\ Rev.\ }\textbf{D10} (1974) 3235.
%
\bibitem{Skyrme:1961vq}
    T.H.R.~Skyrme, A nonlinear field theory,
    \textit{Proc.\ Roy.\ Soc.\ Lond.\ }\textbf{A260} (1961) 127.
%
\bibitem{Zahed:1986qz}
    I.~Zahed and G.E.~Brown, The Skyrme model,
    \textit{Phys.\ Rept.\ }\textbf{142} (1986) 1.
%
\bibitem{Witten:1998zw} 
    E.~Witten, Anti de Sitter space, thermal phase transition, 
    and confinement in gauge theories, 
    \textit{Adv.\ Theor.\ Math.\ Phys.\ }\textbf{2} (1998) 505 
    [arXiv:hep-th/9803131].
%
%

\end{thebibliography}
\end{document}